# Bulk-like Compressibility of the Au–Au Metallic Bond in the Atomically Precise $Au_{25}$ Cluster

Camino Martín-Sánchez[a*], Khadijetou Ahmed Ethmane[a], Nicholas Giamboni[a], Latévi Max Lawson Daku[a], Céline Besnard[b], Thomas Bürgi[a*]

[a] Faculté des Sciences, Département de Chimie Physique, Université de Genève, 30 Quai Ernest-Ansermet, CH-1211 Genève, Switzerland
[b] Laboratory of Crystallography, Université de Genève, 24 Quai Ernest-Ansermet, CH-1211 Genève, Switzerland

*Email: Camino.Martinsanchez@unige.ch
*Email: Thomas.Buergi@unige.ch

Abstract

We present a high-pressure single-crystal X-ray diffraction study of the atomically precise $Au_{25}(PET)_{18}^q$ cluster (q = -1, 0) up to 10 GPa under strictly hydrostatic conditions. Our crystallographic analysis provides direct evidence for the pressure-induced phase transitions previously suggested by spectroscopic studies. Structural refinements reveal that the cluster accommodates compression through the reorganization of the flexible ligand shell and secondary distortions of the staple motifs, while the $Au_{13}$ icosahedral core remains intact. Notably, the internal Au-Au distances exhibit a monotonic contraction that quantitatively mirrors the compressibility of bulk gold. This invariant rigidity at the sub-nanometer scale demonstrates that the fundamental stiffness of the metallic bond is preserved regardless of size. Our findings reconcile previous contradictions in the elasticity of metal nanostructures by isolating the intrinsic mechanical response of the gold kernel from extrinsic structural and experimental artifacts.

## 1 Introduction

Atomically precise monolayer-protected metal clusters (MPCs) exhibit size-dependent optical, chiroptical and electronic properties that make them highly promising candidates for applications ranging from catalysis to biomedicine [1-3]. Among them, the $Au_{25}(SR)_{18}^-$ (SR = thiolate ligand) cluster stands out due to its remarkable stability, well-established synthesis, and thoroughly characterized superatomic electronic structure [4]. Despite the extensive research devoted to its electronic and spectroscopic properties, its mechanical response to external stimuli remains far less explored. This lack of studies arrives from the intrinsic experimental challenges associated with probing mechanical properties at the nanoscale, where applying controlled forces and accurately monitoring the resulting structural deformations is not trivial.

In this context, high-pressure techniques provide a powerful and well-defined route to probe such responses, enabling direct access to structural evolution and compressibility. Building on this capability, pressure has increasingly been employed as an external stimulus to modulate and investigate the physical properties of MPCs. In recent years, compression has been proved to be an effective tool to precisely modulate the properties of metal nanoclusters [5-8]. Beyond property tuning, it can also promote the formation of new metastable phases or novel cluster structures [9,10]. Theoretical studies further predict substantial pressure-driven modifications of bonding motifs, including the emergence of intercluster covalent networks at extreme compression [11]. While these works collectively demonstrate the rich pressure-dependent behavior of MPCs, quantitative structural information under compression remains comparatively scarce.

For the $Au_{25}$ family specifically, Q. Li *et al.* [12] reported an estimation of the bulk modulus for $Au_{25}(PET)_{18}^{-}$ of about 17 GPa. More recently, compression of $Au_{25}(SNap)_{18}$ —which preserves the same $Au_{13}$ core–staple architecture but differs in ligand and counterion composition—was investigated [13], yielding a bulk modulus of approximately 3 GPa. These values reflect the compressibility of the crystalline assembly as a whole, where the higher compressibility of the ligand due to its softer nature significantly contributes to the overall mechanical response. Beyond the global compressibility of the crystal, a fundamental question emerges: how does the $Au_{13}$ core itself respond to external pressure? This question emerges among a long-standing debate regarding whether metallic compressibility remains invariant or diverges at the nanoscale [14-20]. In the case of gold, early studies reported compressibility values varying by as much as 70%, with bulk modulus values ranging from 133 GPa to 290 GPa [21-23], compared to the well-stablished 167 GPa of bulk gold [24]. However, more recent works suggests that both single-crystal [25] and penta-twinned nanoparticles [26] exhibit a compressibility only slightly higher (around 2%) than that of bulk gold. Whether a 13-atom gold kernel —at the limit of miniaturization— retains bulk-like mechanical behavior or instead exhibits different elasticity is still unknown.

Additionally, high-pressure techniques have also been employed to explore pressure-induced phase transitions in $Au_{25}(PET)_{18}^{-}$ [27], where three phase transitions were identified at approximately 0.8, 6 and 11 GPa through optical absorption and Raman spectroscopy. The two first transitions were attributed to a conformational rearrangement within the ligand shell, leaving the metallic core essentially unaffected. However, these interpretations rely on indirect spectroscopic evidence, and direct structural validation remains lacking. How the cluster structure evolves at the atomic level across these phase transitions remains an open question.

In this work we present direct structural evidence of the atomic-scale evolution of $Au_{25}(PET)_{18}^{q}$ (q = -1, 0) under hydrostatic compressions of up to 10 GPa. By performing high-pressure single-crystal X-ray diffraction (SC-XRD), we accurately determine the equation of state (EOS) of the crystalline assembly and track the pressure-dependent evolution of the Au-Au interatomic distances within the core. Our results reveal that the previously reported phase transitions are driven by a hierarchical response to compression: the structural transition is primarily accommodated by the reorganization of the flexible ligand shell, which in turn induces secondary distortions in the staples, while the core remains remarkably robust. Furthermore, we demonstrate that the metallic kernel exhibits a compressibility quantitatively consistent with that of bulk gold. This finding indicates that the fundamental stiffness of the Au–Au metallic bond is preserved even at the sub-nanometer scale, providing a reference point for discussions on size-dependent elasticity in metallic nanomaterials.

# 2 Results and Discussion

## *2.1 Global response under pressure*

Figure 1 shows the variation of the lattice volume with pressure of the $[Au_{25}(PET)_{18}]^{-}[TOA]^{+}$ (referred as $Au_{25}(PET)_{18}^{-}$ from now on) single crystal in the 0-10 GPa pressure range. Our results are in good agreement with the experimental results previously reported by Q. Li *et al*. [12]. Additionally, the different crystals studied in the 4 runs of measurements provide consistent $V(P)$ values. Remarkably, the unit cell volume decreases monotonically over the entire pressure range, with no discernible discontinuities at 0.8 or 6 GPa where pressure-induced phase transitions were previously identified spectroscopically [27]. In this way, the whole experimental $V(P)$ data set in the 0-10 GPa was fitted to a single Vinet EOS [28]

$$P = 3K_0 \frac{1-f}{f^2} \exp\left[\frac{3}{2}(K_0' - 1)(1-f)\right] \quad (1)$$

with $f = (V/V_0)^{1/3}$, where $K_0$ is the bulk modulus, $K_0'$ is the pressure derivative of the bulk modulus, and $V_0$ is the lattice volume at zero pressure. As the mechanical properties of $Au_{25}$ at a quantitative level are poorly addressed in the literature, none of the parameters were fixed during the fitting. The obtained values are $V_0 = 4940.1(1.2)$ Å$^3$, $K_0 = 5.1(2)$ GPa, and $K_0' = 9.2(2)$. The $K_0$, and $K_0'$ confidence ellipse is shown in Figure 1. Additionally, the correlation coefficients between the fitting variables and the fitting variance-covariance matrix are shown in section 1 in the Supporting Information (SI). The obtained bulk modulus confirms that $Au_{25}(PET)_{18}^{-}$ behaves as a highly compressible solid, with a stiffness comparable to that of molecular crystals and more than one order of magnitude lower than bulk gold. This indicates that, indeed, the global mechanical response of the crystalline assembly is dominated by the compression of the ligands rather than by the intrinsic rigidity of the metallic core. Importantly, despite its high compressibility, the initial unit cell volume is fully recovered upon decompression, indicating full reversible structural behavior within the explored pressure range.

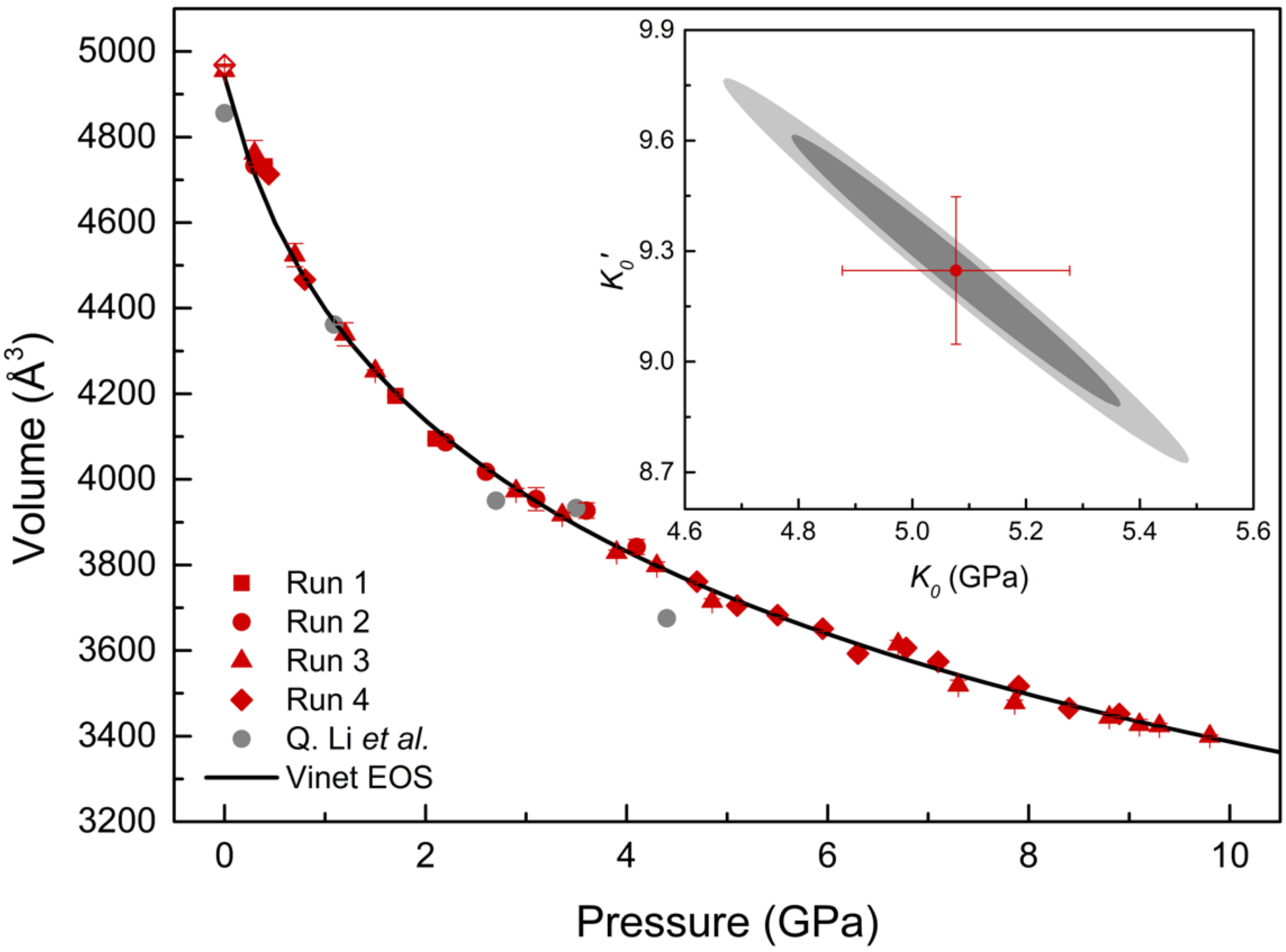

Figure 1. Pressure dependence of the $Au_{25}(PET)_{18}^{-}$ unit cell volume. Filled and empty red symbols correspond to experimental data reported in the present work measured in upstroke and downstroke, respectively; grey symbols correspond to previously reported experimental data by Q. Li *et al.* [12]; solid line corresponds to a fit of the Vinet EOS to the experimental $V(P)$ data. Error bars in volume are either indicated or smaller than the symbols. Inset shows the confidence ellipse for $K_0$, and $K_0'$ at the 68.3% (darker) and 90% (lighter) confidence limits.

The evolution of the individual lattice parameters is presented in Figure 2. The compressibility of the unit cell parameters *a, b,* and *c* is slightly anisotropic. At 10 GPa and in descending order, *a* is compressed by 13%, *c* by 12% and *b* by 11%. No abrupt changes in the lattice dimensions are observed at 0.8 or 6 GPa. However, a closer inspection of the low-pressure region reveals signatures of the first phase transition. Linear equation-of-state fits applied independently to $a(P)$, $b(P)$, and $c(P)$ fail to reproduce the experimental data below the first phase transition pressure, while providing a satisfactory description at higher pressures. Additionally, at approximately 0.6 GPa, $\beta$ decreases sharply from about 106º to 103º, while $\gamma$ simultaneously increases from 91º to 96º. These variations are fully consistent with the ambient-pressure structure previously reported with $\beta = 105.659(7)$ º and $\gamma = 90.859(7)$º [29], and with the compressed structure at 1.1 GPa described by Q. Li *et al.* [12] with $\beta = 102.72(3)$ º and $\gamma = 96.19(2)$º, confirming that the discontinuity in angular parameters corresponds to the first pressure-induced phase transition identified spectroscopically [27].

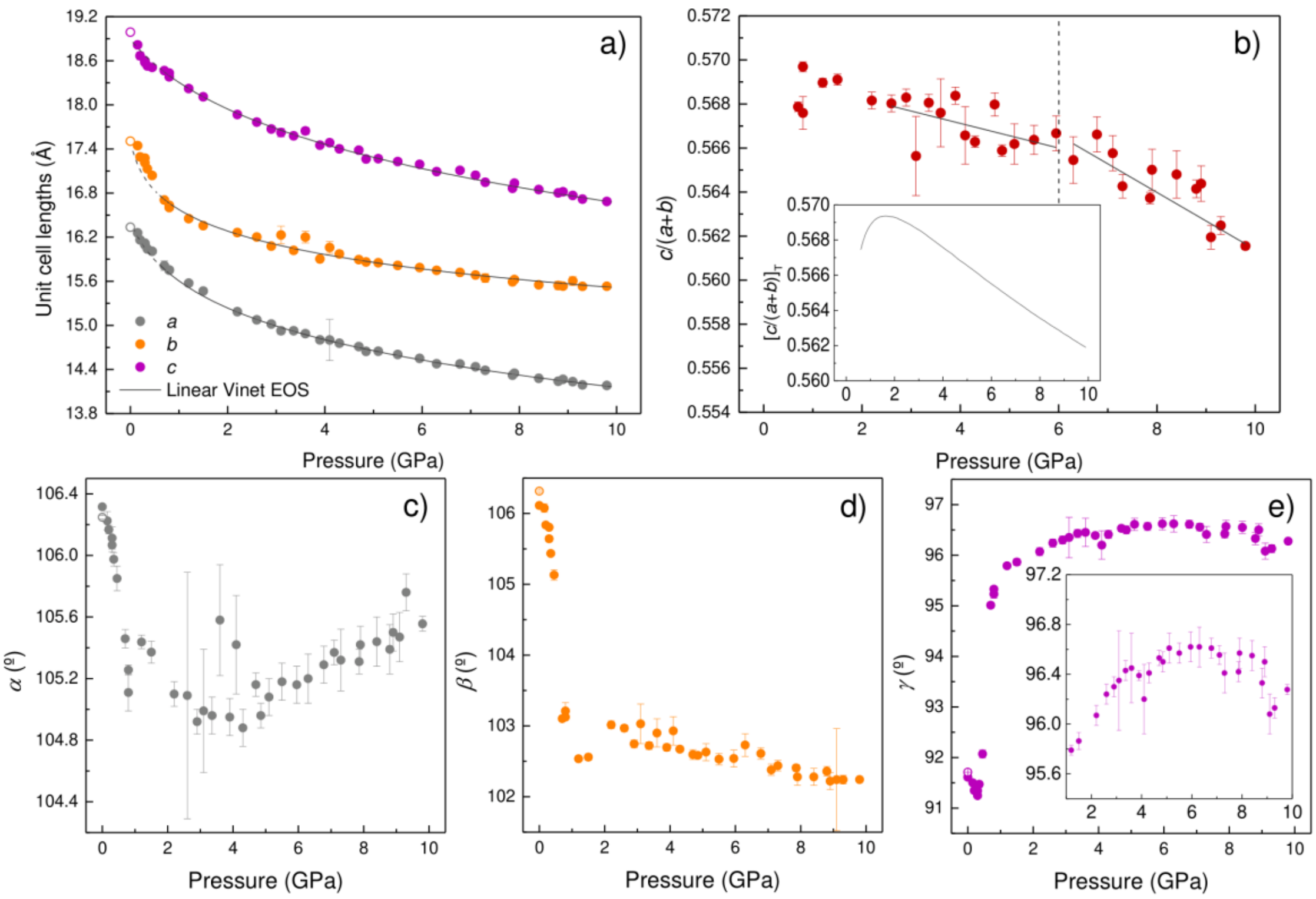

Figure 2. Pressure dependence of the lattice parameters of $Au_{25}(PET)_{18}^{-}$ a) Unit cell axes lengths as a function of pressure. Filled and empty symbols correspond to experimental data measured in upstroke and downstroke, respectively; solid line corresponds to a fit of a linear Vinet EOS to the measured data. The fit below ~0.6 GPa is shown as a dashed line to indicate extrapolation of the EOS to the low-pressure region. b) $c/(a+b)$ ratio as a function of pressure. Filled symbols correspond to experimental data; solid line corresponds to the linear least squares' regression fit to the data. Inset shows the theoretical behavior of the $c/(a+b)$ calculated from the $a(P)$, $b(P)$, and $c(P)$ linear EOS. c), d) and e) $\alpha$, $\beta$, and $\gamma$, respectively, as a function of pressure. The inset in e) shows a magnification of the $\gamma(P)$ data.

In contrast, the second transition at ~6 GPa is subtler and is not reflected in the unit cell volume or lattice lengths. To probe possible weak structural signatures, and in order to cancel any systematic errors we have evaluated the pressure dependence of the $c/(a+b)$ ratio, chosen for its nearly linear behavior in the relevant pressure range. According to the theoretical behavior of the ratio obtained from the calculated linear EOS $a(P)$, $b(P)$ and $c(P)$, $c/(a+b)_T$ shows a continuous linear decrease above 2.5 GPa (see Figure 2b). On the contrary, the experimental data exhibit a change in slope at ~6 GPa. This subtle anomaly, together with the non-monotonic behavior of the $\alpha$ angle, initially decreasing and subsequently increasing beyond 5-6 GPa, provides structural evidence for the second phase transition. Notably, similar lattice parameter ratios have been previously used to identify subtle phase transitions with no clear changes in the volume nor in the unit cell lengths [30]. Comparable situations have been reported in other molecular systems, where clear vibrational discontinuities under pressure reveal structural transitions that remain barely detectable in conventional crystallographic metrics [31,32].

### *2.2 High-pressure structure*

While the unit cell and lattice parameter analysis clearly confirm the presence of pressure-induced phase transitions, the specific changes occurring within the cluster structure itself remain unresolved. To elucidate how these transitions affect the cluster architecture, we acquired high-quality single-crystal data on the neutral species, $Au_{25}(PET)_{18}^0$, in order to eliminate the counter-ion and allow a more reliable refinement of the structure. The neutral and anionic clusters share the same core and staple framework [29, 33], while differences between them mainly arise from distortions of the ligand shell induced by the presence of the counterion [34], enabling direct comparison of the intrinsic cluster structure.

To validate the use of the neutral species as a structural analogue for the anionic cluster under compression, we first characterized its global response. The pressure dependence of the unit cell parameters for $Au_{25}(PET)_{18}^0$ (see Figure S17 in the SI) exhibits equivalent structural signatures previously identified for the anionic species. The first phase transition is evidenced by a singular expansion of the $a$-axis accompanied by a rapid contraction of the $c$-axis, alongside sharp discontinuities in $\alpha$ and $\gamma$. Since the structure is triclinic, with the central atom of the cluster core lying at the center of the unit cell, the intercluster distances correspond directly to the unit-cell parameters. Consequently, the pronounced initial contraction along the $c$-axis can be directly attributed to its comparatively larger initial separation, whereas the already shorter $a$-axis exhibits only limited compression at the beginning of the pressurization (see Figure S17). Beyond this regime, the system compresses more uniformly until approximately 5 GPa, where the second transition occurs. This transition takes place at slightly lower pressures compared to the anionic species (~6 GPa), a trend consistent with previous findings in neutral $Au_{25}$ dimers [27]. This is most clearly manifested in the pressure dependence of the $\beta$ angle, which initially increases with pressure before reaching a plateau. A comparative analysis of the $V(P)$ data (see Figure S18 in the SI) reveals that the relative volume contraction of the neutral species is essentially identical to that of its anionic counterpart, leading to a comparable bulk modulus (ca. 5 GPa). These results show that the neutral and the anionic clusters show similar pressure dependent structure changes.

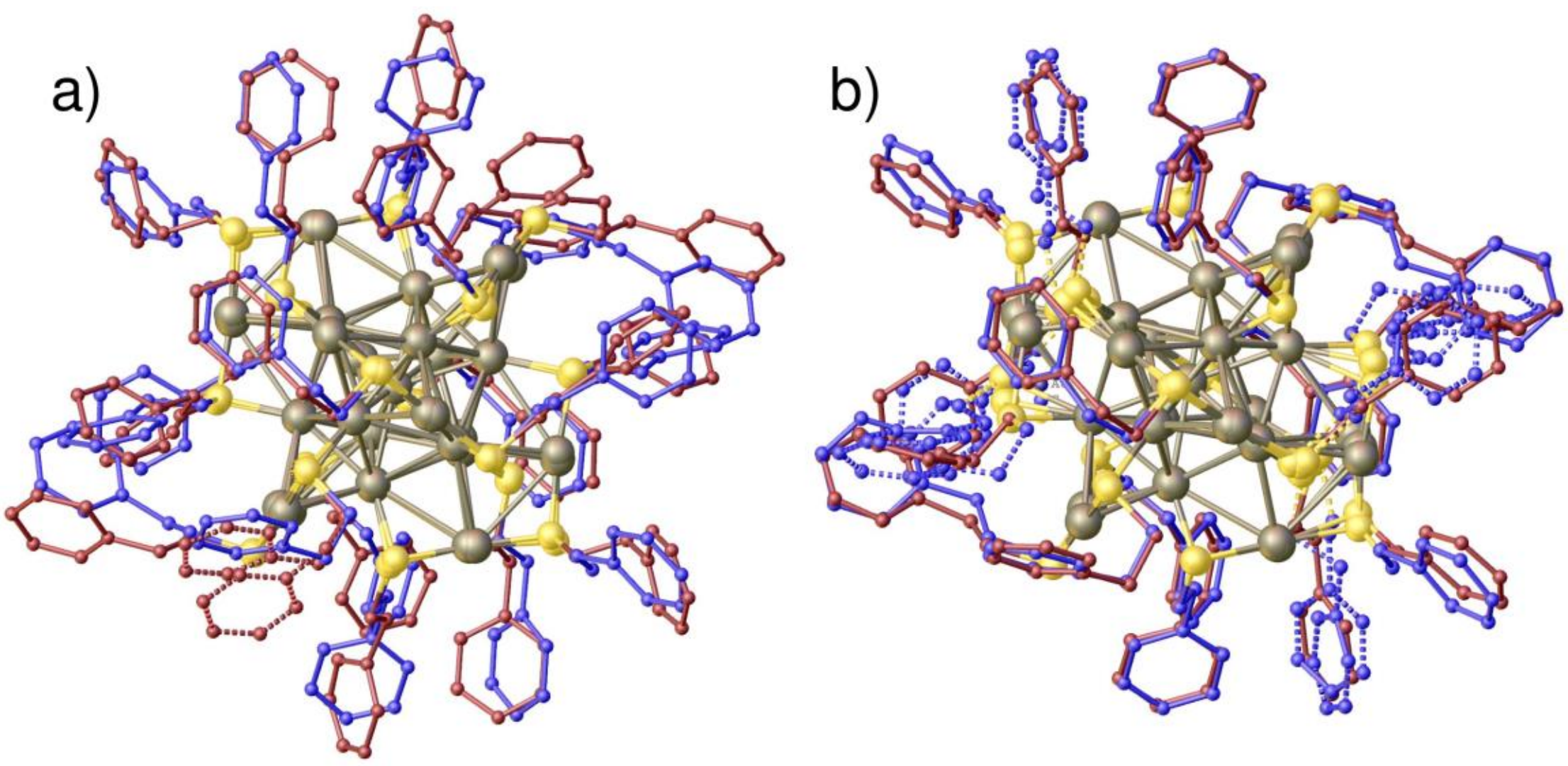

Figure 3. Evolution of the structure of $Au_{25}(PET)_{18}^0$ through the phase transitions. Superimposition of the structures before and after the transitions at a) 0.4 GPa and 1.2 GPa, and b) 3.6 GPa and 4.9 GPa. In both panels, carbon atoms of the ligand shell at the lower pressure are shown in red, while those at the higher pressure are in blue. Gold and sulfur atoms are shown in gold and yellow, respectively. Disordered parts are indicated with stippled bonds. All structures were aligned using the gold atoms of the icosahedral $Au_{13}$ core to highlight the ligand reorganization.

Figure 3 shows the evolution of the structure of $Au_{25}(PET)_{18}^{0}$ through the phase transition pressures. The centrosymmetric model of the ambient-pressure structure was maintained, using a restrained model for the ligands (see section 2 in the SI), we successfully refined the structure at the different experimental pressures. Periodic DFT geometry optimizations were performed with no symmetry constraints to check if the structure remains centrosymmetric under pressure. Centrosymmetry was predominantly found in all calculated structures, with a probability of 91% at 4 GPa and still 74% at 10 GPa, supporting the validity of the model (see details in section 6 in the SI). Although the low completeness of the X-ray data (less than 30%) hampered a highly precise structural description, the position of the ligands was confidently confirmed in the Fourier difference maps, and both phase transitions were clearly identified (see Figures 3 and S22). These phase transitions induce a structural response that highlights the different mechanical roles of the cluster's components, where the rigid core is surrounded by a flexible ligand shell that can easily rearrange. In addition, the presence of the phenyl rings enables a rich network of intermolecular interactions—including phenyl–phenyl and S–H interactions—which stabilize the crystal packing. A detailed analysis of the first transition shows that, as pressure increases, the central $Au_{13}$ core undergoes a purely isotropic contraction, preserving its symmetry and structural integrity without any detectable distortion. Meanwhile, the staples undergo a subtle distortion to accommodate the transition: the Au–S bonds within the staples expand by 0.6% and the $Au_{core}$-$Au_{staple}$ distances increase by 0.5%. This expansion is non-homogeneous (see Figures S20 and S23) and is counterbalanced by a 0.5% contraction of the $Au_{core}$-$S_{staple}$ bonds, alongside a pronounced rearrangement of the surrounding ligands (see Figures 3 and S23). This structural reorganization substantially modifies the intercluster interactions (see Figures S24 and S25) as the orientations of some phenyl rings are altered.

Following the first transition, the crystal experienced minor fracturing, yet the diffraction quality remained sufficient to track the structure up to approximately 10 GPa. Beyond 4 GPa a further rearrangement of the ligands occurs, requiring a disordered model with two possible conformations for ligands 2, 3, and 5 (see Figures S22 and S23). Overall, these observations demonstrate that the ligands act as the main adaptive component of the structure, accommodating pressure-induced stresses while the metallic core is comparatively rigid.

### *2.3 Compressibility of the $Au_{13}$ kernel*

To quantitatively assess the mechanical response of the $Au_{13}$ kernel, we investigated its intrinsic compressibility. To accurately quantify the core compressibility, we monitored the pressure evolution of the Au–Au interatomic distances within the $Au_{13}$ icosahedron. Due to the centrosymmetric nature of the cluster, the 12 radial distances between the central gold atom ($Au_c$) and the gold atoms in the icosahedral vertex ($Au_{ic}$), as well as the 30 tangential $Au_{ic}$-$Au_{ic}$ contacts, are defined by 6 and 15 independent crystallographic vectors, respectively.

The icosahedral core exhibits intrinsic structural irregularities, *e.g.* vertex-vertex distances ranging from approximately 2.8 to 3.1 Å at ambient pressure (see Figure S19). This dispersion, primarily originating from Jahn–Teller distortions [34], remains remarkably stable across the entire pressure range studied. Despite this overall stability, a detailed analysis of individual bonds is technically challenging because the atomic positions are correlated during structural refinement. To overcome this, we employed the mean interatomic distance for both radial and vertex-vertex distances as the representative structural observable. By averaging over the entire set of symmetry-independent distances, we effectively neutralize these fluctuations, yielding a monotonic and physically consistent description of the core's contraction (see Figure S19). While this structural irregularity results in a relatively large standard deviation for the absolute average values, it is important to note that this reflects the intrinsic dispersion of the bond distance distributions rather than experimental uncertainty. For the mean distances, the reported experimental uncertainty corresponds to the error derived from the structural refinement, calculated via standard error propagation of the individual bond uncertainties. This value strictly represents the measurement precision and excludes the intrinsic dispersion arising from the geometrical distortion of the core.

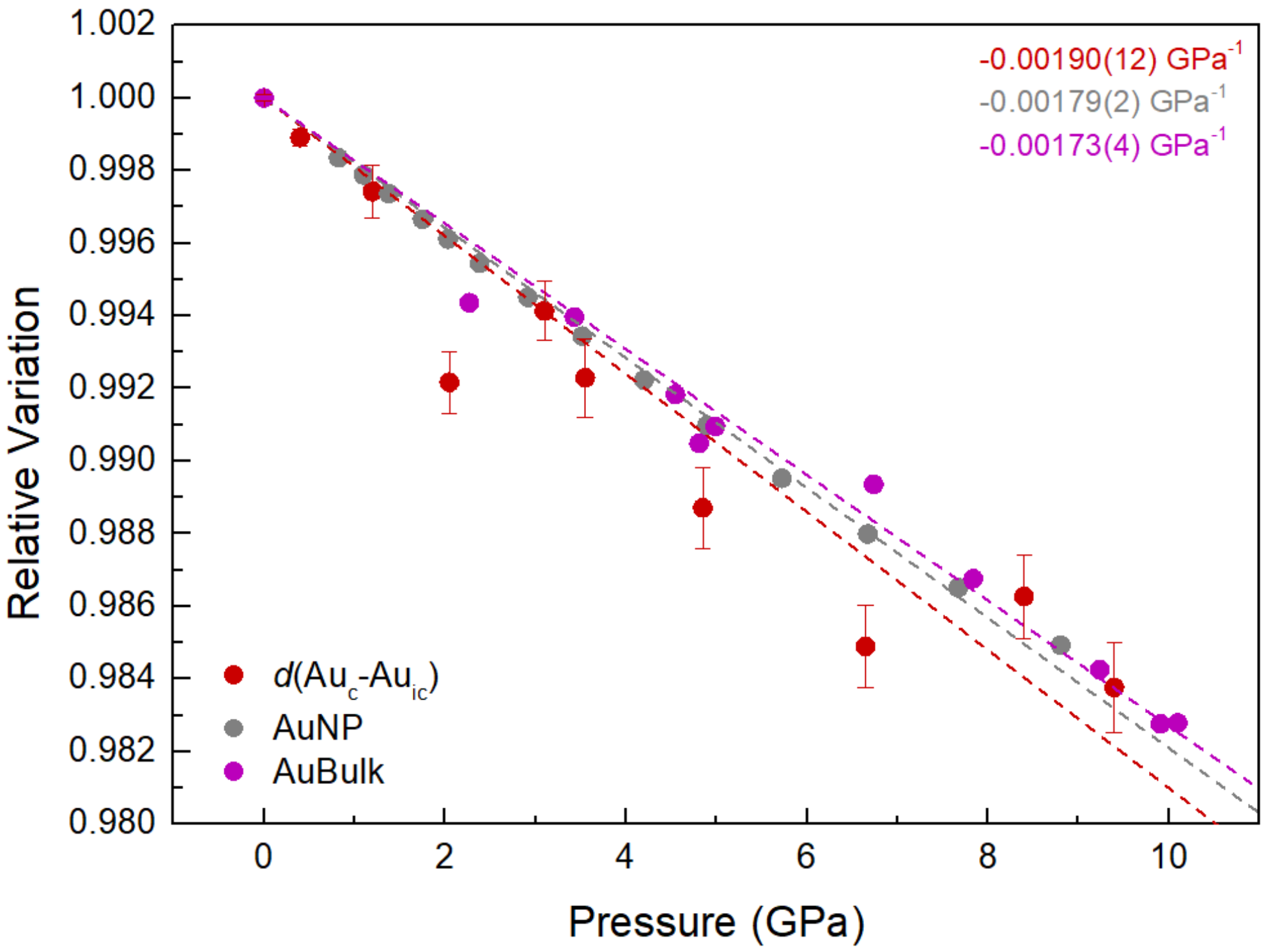

Figure 4. Pressure dependence of the relative variation of the mean radial distances within the $Au_{13}$. For comparison, the relative variation of the lattice parameter of single-crystal gold nanoparticles [25], and bulk gold [24] is included. Dashed lines represent linear fits to the data with the intercept fixed at unity.

Figure 4 presents the pressure evolution of the normalized mean radial $Au_c$-$Au_{ic}$ distance, which characterize the core contraction. Notably, both the radial and the edge $Au_{ic}$-$Au_{ic}$ distances compress at the same rate, reflecting a highly isotropic compression of the core (see Figure S21 in the Supporting Information). For the main discussion, the radial distance was selected as the primary structural descriptor because it exhibits a significantly lower associated experimental uncertainty. For comparison, the compression rates of both bulk gold [24] and single-crystal gold nanoparticles (SC-AuNP) [25] are also displayed, with the core contraction closely mirroring their behavior. To quantify the contraction rates, a linear regression analysis was performed with the intercept fixed at $d/d_0 = 1$ to maintain physical consistency at ambient pressure. The resulting linear contraction coefficient for the $Au_c$-$Au_{ic}$ distance is -0.00190(12) $GPa^{-1}$. It is worth noting here that tracking internal interatomic distances in molecular clusters inherently yields higher data dispersion than standard XRD methods used to extract bulk lattice parameters. Nevertheless, both bulk gold and SC-AuNPs, which exhibit an equivalent bulk modulus ($K_0$ = 167 and 170 GPa, respectively), present a linear compression rate that closely matches this value. Within the experimental uncertainty margin, the contraction rate of the cluster core is statistically compatible with these reference values. Consequently, within the limits of our experimental accuracy, the contraction of the internal Au-Au bonds in the $Au_{25}$ cluster remains consistent with the lattice evolution of larger metallic gold structures. Such agreement provides evidence that the fundamental mechanical compressibility of the gold kernel is preserved, regardless of the drastic reduction in size.

The observed invariance in compressibility is particularly significant given the long-standing discussion regarding size-dependent elastic properties in metallic nanoparticles. While several studies on nanoscaled metals have reported complex variations in bulk moduli [14-18], these discrepancies often stem from a conceptual overlap between the behavior of isolated SC-NP and the grain-boundary-dominated response of nanocrystalline aggregates. In the latter, the global mechanical response is frequently governed by extrinsic factors, such as internal twinning, stacking faults, or high-energy grain boundaries, which obscure the intrinsic elasticity of the metal–metal bond. Furthermore, experimental challenges in high-pressure crystallography, particularly the presence of non-hydrostatic stress, must be considered. Deviatoric stresses in granular or poorly mediated samples are known to introduce significant artifacts, often leading to an overestimation of the bulk modulus [21,35]. In this context, the $Au_{25}$ cluster serves as an atomically precise model that effectively isolates the intrinsic elastic response of the Au–Au metallic bond. Our study, conducted under strictly hydrostatic conditions, allows us to decouple the metal's response from the experimental and structural noise typical of larger nanocrystalline samples. The fact that the $Au_{13}$ kernel undergoes a contraction quantitatively consistent with the bulk gold phase suggests that the fundamental compressibility of gold remains invariant even at the sub-nanometer scale.

A further analysis of the compression of the structure reveals a clear differentiation in the Au-Au compressibility within the $Au_{25}$ cluster. While the internal Au-Au bonds of the core exhibit a contraction of about 2% in the explored pressure range, the $Au_{core}$-$Au_{staple}$ distances decrease at a significantly higher rate of 5% (see Figure S20). This disparity further highlights that the mechanical response is not uniform across the structure. The $Au_{13}$ kernel preserves the stiff, metallic-like character of bulk gold, whereas the peripheral gold atoms —which are in a different chemical environment and coordinated with sulfur—respond more readily to external pressure. This distinct compression rate is consistent with the structural evolution predicted by DFT simulations [11], which describe a rapid tightening of the staples around the kernel. These computational results associate the higher contraction of the $Au_{core}$-$Au_{staples}$ distance with the closure of the free volume between the kernel and the protecting units, which at higher pressures leads to an increased coordination and the formation of new Au-Au and Au-S contacts.

# 3 Conclusions

In summary, a systematic high-pressure single-crystal X-ray diffraction study has been conducted on the $Au_{25}(PET)_{18}^{q}$ cluster (q = -1, 0) to unravel its mechanical and structural response upon compression. Our structural analysis reveals that the two-phase transitions that the cluster undergoes are manifested through significant reorganizations of the ligand shell which in turn distort the staple framework. Noteworthily, the $Au_{13}$ icosahedral core remains remarkably stable through these phase transitions, preserving its structural integrity across the entire explored pressure range.

By disentangling the intrinsic response of the metallic kernel from the ligand shell's high compressibility, we demonstrate that the $Au_{13}$ core exhibits a linear contraction coefficient consistent with the compressibility of bulk gold, providing a definitive answer to the long-standing debate regarding the mechanical behavior of metals at the sub-nanometer scale. Our results indicate that at the limit of miniaturization, the Au-Au metallic bond retains the fundamental elastic properties of the macroscopic metal.

More broadly, our findings demonstrate that the structural plasticity of these clusters enables ligand reorganization at accessible pressures, opening a path toward new phases with tailored properties. We hope this work stimulates future research to explore the possibilities of high pressure and the inherent versatility of ligand functionalization to unlock novel properties in atomically precise metal nanoclusters.

# 4 Experimental section

## 4.1 Single-crystal synthesis

*Synthesis and crystallization of $[Au_{25}(PET)_{18}]^{-}[TOA]^{+}$*

$Au_{25}(PET)_{18}^{-}$ was synthesized based on the procedure previously reported by Kalenius *et al.* [36], with a few modifications.

First, 980 mg of TOABr (1.79 mmol) were dissolved in 100 mL THF before the addition of 1 g $HAuCl_4 \cdot 3H_2O$ (2.52 mmol), under stirring. An additional 150 mL of THF was then poured into the reaction mixture which was subsequently stirred for 30 min at room temperature. After that, 1.7 mL of PET (12.7 mmol) was added dropwise and the solution was left agitating for 50 min. Finally, a pre-cooled solution of 965 mg $NaBH_4$ (25.5 mmol) in 50 mL MilliQ water (> 18 MΩ cm) was rapidly added to the solution. After 32 hours of stirring, the resulting product was concentrated using a rotavapor and the aqueous phase was removed. Methanol was added and left overnight for the excess thiols and water to transfer to the alcohol while $Au_{25}(PET)_{18}^{-}$ precipitates. The precipitate was then isolated by filtration and dissolved in DCM before it was dried using a rotavapor.

$Au_{25}(PET)_{18}^{-}$ was then dissolved in toluene and single-crystals were obtained by diffusion with ethanol after two weeks at room temperature.

*Synthesis and crystallization of $Au_{25}(PET)_{18}^{0}$*

Neutral $Au_{25}(PET)_{18}^{0}$ is obtained by oxidation of $Au_{25}(PET)_{18}^{-}$ [34]. The negatively charged species was solubilized in a minimum amount of toluene and run through a silica column, with a 1:1 solution of DCM:hexane as the mobile phase. The collected product was then dried and re-solubilized in toluene. Single-crystals are grown by diffusion of ethanol in toluene after two weeks at 4°C.

## 4.2 High-pressure X-ray diffraction measurements

High-pressure experiments were performed in an Almax easyLab Diacell® Bragg-Mini diamond anvil cell equipped with 500 μm anvil culet diameters. The 200 μm thick Inconel 301 gaskets were preindented to 40-60 μm. The cylindrical pressure chamber was made by perforating a 200 μm diameter hole in the center of the preindented gasket using a BETSA motorized electric discharge machine. Methanol-ethanol (MeOH-EtOH) 4:1 mixture was used as pressure transmitting media (PTM) as it solidifies at around 11 GPa [37], providing a relatively wide hydrostatic pressure range for the precise determination of $V(P)$ with a precision of $\Delta V/V \sim 10^{-3}$. The hydrostaticity of the PTM was monitored through the pressure dependence of the full width at half-maximum of the ruby emission R1 line [38], whereas the pressure inside the cavity was determined from the ruby R1-line shift following the recently updated pressure scale [39].

Single-crystal X-ray diffraction data were acquired on a commercial Synergy – DW rotating anode (Rigaku) diffractometer, equipped with a Mo-Kα microfocus source (λ = 0.7093 Å) and a HyPix-Arc 150º curved photon-counting detector. Full datasets suitable for structure refinement of $Au_{25}(PET)_{18}^{0}$ required long measurement times (~5h per measurement), whereas shorter acquisitions were performed to determine lattice parameters of $Au_{25}(PET)_{18}^{-}$ across the full pressure range. To ensure accurate and reproducible determination of the unit cell volume as a function of pressure, four crystals were measured in four independent runs. The resulting $V(P)$ data showed full reproducibility between runs, confirming the reliability of the measurements for subsequent EOS analysis. The fitting of the $V(P)$ data to a Vinet EOS was made through EosFit7-GUI [40].

## Data availability statement

The crystallographic datasets generated during the current study are available in the Zenodo repository with the identifier https://doi.org/10.5281/zenodo.20590917.

## Supporting Information

Supporting Information contains detailed information on the following items: (1) Lattice Parameters, Equation of State Parameters and Statistical Refinement Data for $Au_{25}(PET)_{18}^{-}$; (2) Structural refinement model and crystallographic data for $Au_{25}(PET)_{18}^{0}$; (3) Pressure dependence of the lattice parameters of $Au_{25}(PET)_{18}^{0}$; (4) Pressure dependence of the atomic distances within $Au_{25}(PET)_{18}^{0}$; (5) Pressure-Induced Structural Rearrangements and Ligand Shell Conformational Evolution of $Au_{25}(PET)_{18}^{0}$; (6) Computational Assessment of Structural Centrosymmetry of $Au_{25}(PET)_{18}^{0}$ under Pressure.

## Acknowledgements

T. B. acknowledges the generous support from the Swiss National Science Foundation (grant 200020_214996). We thank Dr. Rosario Scopelliti and Dr. Pascal Schouwink from the X-Ray Diffraction and Surface Analytics Platform of the EPFL for support in the X-ray diffraction measurements.

## Abbreviations

DAC, diamond anvil cell; DCM, dichloromethane; EOS, equation of state; MeOH-EtOH, methanol-ethanol; MPCs, monolayer-protected metal nanoclusters; PET, 2-phenylethanethiol; PTM, pressure-transmitting medium; SC-AuNP, single-crystal gold nanoparticles; THF, tetrahydrofuran; TOA, tetraoctylammonium; TOABr, tetraoctylammonium bromide; XRD, X-ray diffraction.

# Supporting Information

## 1) Lattice Parameters, Equation of State Parameters and Statistical Refinement Data for $Au_{25}(PET)_{18}^{-}$

Table S1. Lattice parameters as a function of pressure

| Pressure (GPa) | *a* (Å) | *b* (Å) | *c* (Å) | $\alpha$ (º) | $\beta$ (º) | $\gamma$ (º) | *V* (Å$^3$) |
|---|---|---|---|---|---|---|---|
| Run 1 | | | | | | | |
| 0.3 | 16.0652(68) | 17.21(2) | 18.577(8) | 106.113(72) | 105.64(4) | 91.343(64) | 4734(6) |
| 2.2 | 15.186(8) | 16.263(24) | 17.868(12) | 105.10(8) | 103.015(45) | 96.07(8) | 4087(6) |
| 2.6 | 15.0748(76) | 16.202(24) | 17.766(12) | 105.1(8) | 102.968(48) | 96.24(8) | 4018(6) |
| 3.1 | 14.926(48) | 16.23(12) | 17.623(56) | 105.0(4) | 103.03(28) | 96.4(4) | 3954(27) |
| 3.6 | 14.887(32) | 16.20(8) | 17.645(48) | 105.58(36) | 102.9(2) | 96.45(28) | 3927(18) |
| 4.1 | 14.801(28) | 16.06(8) | 17.49(4) | 105.42(32) | 102.9(2) | 96.20(28) | 3842(18) |
| Run 2 | | | | | | | |
| 0.3 | 16.119(6) | 17.276(12) | 18.5996(76) | 106.067(44) | 105.804(36) | 91.25(4) | 4762(30) |
| 1.2 | 15.5736(52) | 16.452(12) | 18.2215(64) | 105.438(44) | 102.535(28) | 95.79(4) | 4339(27) |
| 1.5 | 15.4658(76) | 16.356(16) | 18.110(8) | 105.372(72) | 102.559(44) | 95.864(68) | 4253(3) |
| 2.9 | 15.017(8) | 16.078(24) | 17.671(12) | 104.92(8) | 102.476(56) | 96.30(8) | 3973(6) |
| 3.4 | 14.926(8) | 16.020(28) | 17.579(12) | 104.96(12) | 102.721(52) | 96.46(8) | 3917(6) |
| 3.9 | 14.802(4) | 15.903(32) | 17.452(12) | 104.95(12) | 102.697(52) | 96.39(4) | 3829(6) |
| 4.3 | 14.7563(64) | 15.972(44) | 17.401(8) | 104.88(12) | 102.67(4) | 96.41(8) | 3798(9) |
| 4.9 | 14.6459(64) | 15.862(32) | 17.264(8) | 104.96(8) | 102.58(4) | 96.50(8) | 3715(6) |
| 7.3 | 14.388(12) | 16.64(6) | 16.946(16) | 105.3(2) | 102.438(76) | 96.41(16) | 3519(12) |
| 7.9 | 14.322(8) | 15.593(32) | 16.864(8) | 105.31(8) | 102.407(48) | 96.42(8) | 3478(6) |
| 8.8 | 14.240(12) | 15.543(56) | 16.802(12) | 105.39(16) | 102.359(68) | 96.33(12) | 3444(9) |
| 9.1 | 14.233(12) | 15.60(6) | 16.767(16) | 105.47(16) | 102.24(72) | 96.08(16) | 3427(12) |
| 9.3 | 14.190(8) | 15.530(36) | 16.717(12) | 105.76(12) | 102.24(6) | 96.13(8) | 3424(6) |
| 9.8 | 14.1816(32) | 15.532(16) | 16.6857(44) | 105.556(48) | 102.24(2) | 96.28(4) | 3400(3) |
| Run 3 | | | | | | | |
| 0.8 | 15.752(16) | 16.63(2) | 18.381(24) | 105.11(12) | 103.21(12) | 95.23(8) | 4467(6) |
| 4.7 | 14.711(8) | 15.894(12) | 17.383(16) | 105.161(76) | 102.595(65) | 96.53(6) | 3461(3) |
| 5.1 | 16.647(16) | 15.852(24) | 17.268(28) | 105.08(12) | 102.63(12) | 96.61(12) | 3705(6) |
| 5.5 | 14.604(16) | 15.82(2) | 17.23(2) | 105.18(12) | 102.53(8) | 96.57(8) | 3683(6) |
| 6.0 | 14.551(16) | 15.79(2) | 17.191(24) | 105.16(12) | 102.54(12) | 96.62(12) | 3651(6) |
| 6.3 | 14.478(24) | 15.749(32) | 17.092(24) | 105.20(16) | 102.73(16) | 96.62(16) | 3593(9) |
| 6.8 | 14.475(16) | 15.72(2) | 17.109(24) | 105.29(12) | 102.61(8) | 96.61(8) | 3606(6) |
| 7.1 | 14.437(12) | 15.685(16) | 17.042(24) | 105.37(8) | 102.38(8) | 96.555(76) | 3574(6) |
| 7.9 | 14.35(2) | 15.622(24) | 16.934(28) | 105.42(12) | 102.28(12) | 96.57(12) | 3517(6) |
| 8.4 | 14.28(2) | 15.551(28) | 16.848(32) | 105.44(16) | 102.28(12) | 96.55(12) | 3465(6) |
| 8.9 | 14.265(16) | 15.53(2) | 16.818(24) | 105.50(12) | 102.22(12) | 96.50(12) | 3452(6) |
| Run 4 | | | | | | | |
| 0.1 | 16.2578(68) | 17.4451(38) | 18.817(16) | 106.23(6) | 106.08(6) | 91.499(32) | 4893(10) |
| 0.2 | 16.1633(36) | 17.2820(68) | 18.668(52) | 106.168(28) | 105.837(24) | 91.346(24) | 4793(3) |
| 0.3 | 16.040(32) | 17.130(12) | 18.5302(48) | 105.975(44) | 105.435(2) | 91.47(4) | 4691(30) |
| 0.5 | 16.008(8) | 17.039(16) | 18.51(2) | 105.85(8) | 105.131(729 | 92.07(8) | 4641(30) |
| 0.7 | 15.813(64) | 16.705(16) | 18.4661(72) | 105.46(6) | 103.102(32) | 95.012(56) | 4521(3) |
| 0.8 | 15.7529(48) | 16.6017(52) | 18.4321(72) | 105.256(32) | 103.124(28) | 95.328(24) | 4469(18) |

Table S2. Correlation coefficients in % between the fitting variables.

| Variable | $V_0$ | $K_0$ | $K_0'$ |
|---|---|---|---|
| $V_0$ | 100.0 | -8.6 | 3.7 |
| $K_0$ | -8.6 | 100.0 | -98.9 |
| $K_0'$ | 3.7 | -98.9 | 100.0 |

Table S3. Variance-covariance matrix from the least-squares fitting.

| Variable | $V_0$ | $K_0$ | $K_0'$ |
|---|---|---|---|
| $V_0$ | 1.47389 | -0.01970 | 0.01078 |
| $K_0$ | -0.01970 | 0.03578 | -0.04451 |
| $K_0'$ | 0.01078 | -0.04451 | 0.05666 |

Table S4. Vinet equation of state parameters for the unit cell lengths $a$, $b$, and $c$, where $M_0$ and $M_0'$ are the linear modulus and its pressure derivative, respectively.

| Axis length | $x_0$ (Å) | $M_0$ (GPa) | $M_0'$ |
|---|---|---|---|
| *a* | 16.43(3) | 9.4(1) | 28(6) |
| *b* | 17.52(1) | 5(2) | 49(8) |
| *c* | 18.89(9) | 23(4) | 21.1(1.7) |

## 2) Structural refinement model and crystallographic data for $Au_{25}(PET)_{18}^0$

# The refinement model

To overcome the low completeness of the data, a heavily restrained model was made. All ligands were refined as residues, sharing the following restraints:

**Ligand:**

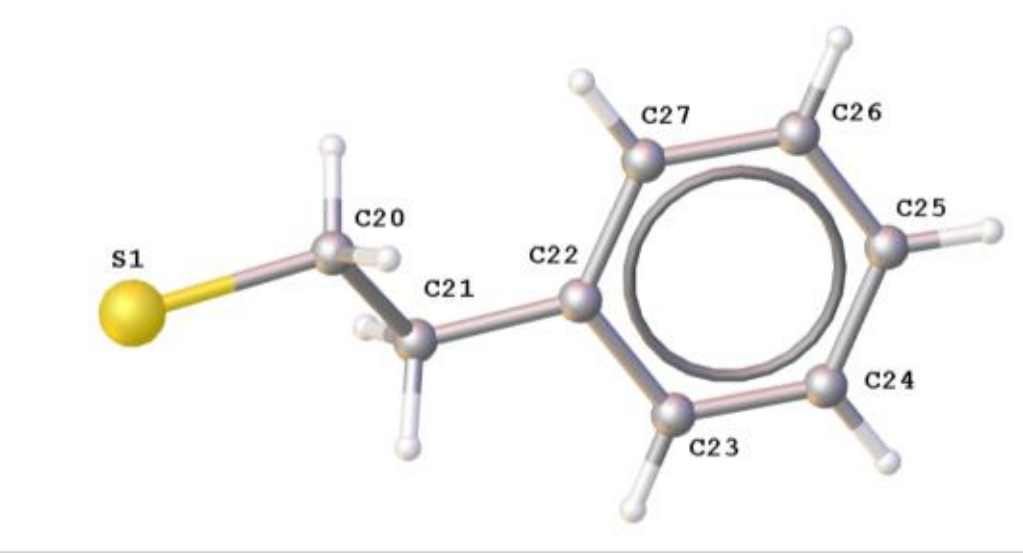

Figure S1. Numbering for the ligand residues.

All 1-2 distances were restrained:
DFIX_LIG 1.83 S1 C20
DFIX_LIG 1.51 C21 C20 C21 C22
DFIX_LIG 1.39 C23 C22 C24 C23 C25 C24 C26 C25 C27 C26 C27 C22
All 1-3 distances were restrained:
DANG_LIG 2.77 S1 C21
DANG_LIG 2.52 C20 C22 C21 C27 C21 C23
DANG_LIG 2.40 C22 C24 C23 C25 C24 C26 C25 C27 C26 C22 C27 C23
1-4 distances in the aryl were also restrained:
DANG_LIG 2.77 C22 C25 C27 C24 C23 C26

The aryl was restrained to be flat:
FLAT_LIG C22 C23 C24 C25 C26 C27 C21

Displacement parameters were also restrained:
RIGU_LIG 0.01 0.02 S1 C20 C21 C22 C27 C26 C25 C24 C23

**Solvent toluene molecule:**

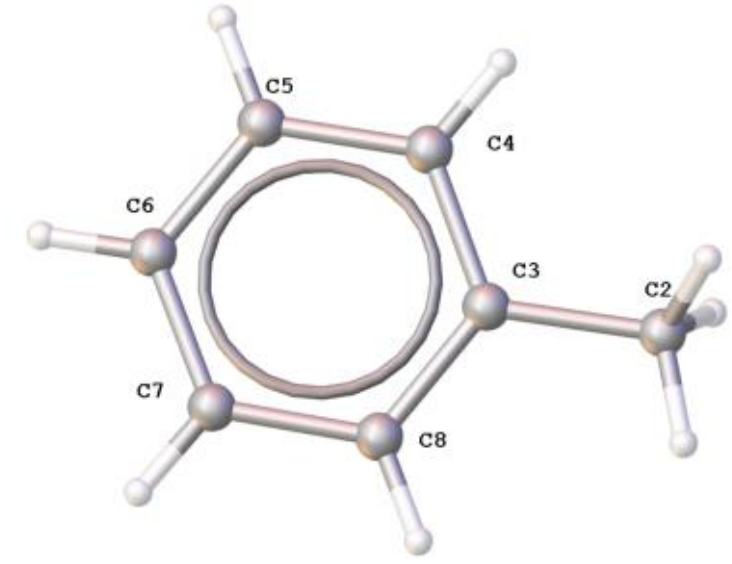

Figure S2. Numbering for the toluene molecule.

DFIX 1.51 C2 C3
DANG 2.52 C2 C4 C2 C8
DFIX 1.39 C3 C4 C3 C8 C5 C6 C4 C5 C6 C7 C7 C8
FLAT C2 C3 C4 C5 C6 C7 C8
DANG 2.40 C3 C5 C4 C6 C5 C7 C6 C8 C7 C3 C8 C4
DANG_LIG 2.77 C3 C6 C4 C7 C5 C8
SIMU C2 C3 C4 C8 C5 C7 C6

Moreover, one of the ligands was disordered over 2 positions and was refined using 2 components. (lig 9 and 10). The following constrained were added:
EADP S1_9 S1_10
EADP C20_9 C20_10
EXYZ S1_9 S1_10

# The sequential refinement

Indexation was made using similar orientation matrices, so that the axes corresponded. As the result the refined model at pressure n-1 was used as a starting model at pressure n.

**P00 FINAL MODEL**

- **Crystal data and structure refinement for P00, 0 GPa.**

| | |
|---|---|
| Identification code | P00 |
| Empirical formula | $C_{158}H_{178}Au_{25}S_{18}$ |
| Formula weight | 7578.24 |
| Temperature/K | 292.23(10) |
| Crystal system | triclinic |
| Space group | P-1 |
| a/Å | 16.2101(4) |
| b/Å | 17.8749(4) |
| c/Å | 18.2797(4) |
| α/° | 65.026(2) |
| β/° | 64.450(2) |
| γ/° | 80.877(2) |
| Volume/Å$^3$ | 4330.1(2) |
| Z | 1 |
| $\rho_{calc}$g/cm$^3$ | 2.906 |
| μ/mm$^{-1}$ | 41.022 |
| F(000) | 3389.0 |
| Crystal size/mm$^3$ | 0.145 × 0.122 × 0.009 |
| Radiation | Cu Kα (λ = 1.54184) |
| 2Θ range for data collection/° | 5.456 to 152.202 |
| Index ranges | -18 ≤ h ≤ 20, -22 ≤ k ≤ 22, -22 ≤ l ≤ 22 |
| Reflections collected | 68913 |
| Independent reflections | 17090 [$R_{int}$ = 0.0531, $R_{sigma}$ = 0.0498] |
| Data/restraints/parameters | 17090/846/974 |
| Goodness-of-fit on F$^2$ | 0.983 |
| Final R indexes [I>=2σ (I)] | $R_1$ = 0.0444, w$R_2$ = 0.1137 |
| Final R indexes [all data] | $R_1$ = 0.0543, w$R_2$ = 0.1207 |
| Largest diff. peak/hole / e Å$^{-3}$ | 1.61/-2.33 |

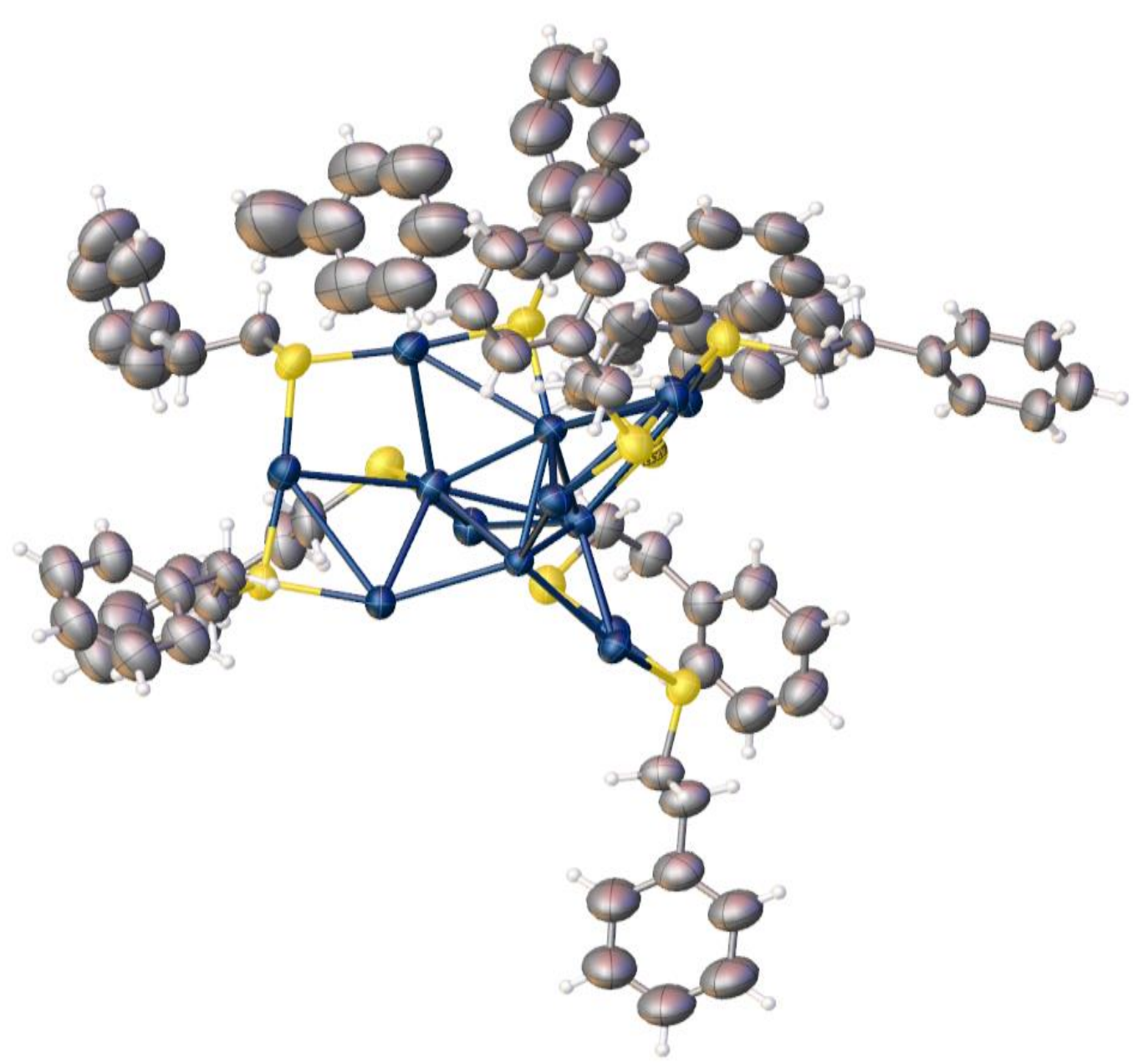

Figure S3. View of the asymmetric unit at 0 GPa along $\vec{a}$ with displacement parameters at 50 percent probability level.

- **Checkcif alerts and answers**

# start Validation Reply Form
_vrf_PLAT342_p00
;
PROBLEM: Low Bond Precision on  C-C Bonds ...............    0.02406 Ang.
RESPONSE: The diffraction data are of good quality but the ligand positions are less well defined as can be shown by the larger ellipsoids.
# end Validation Reply Form

**FROM P00 TO P01**

The Au and S atoms were refined anisotropically and the C atoms isotropically.

**P01 FINAL MODEL**

- **Crystal data and structure refinement for P01, 0.4 GPa.**

| | |
|---|---|
| Identification code | P01 |
| Empirical formula | $C_{158}H_{178}Au_{25}S_{18}$ |
| Formula weight | 7578.24 |
| Temperature/K | 298.6(2) |
| Crystal system | triclinic |
| Space group | P-1 |
| a/Å | 16.0171(7) |
| b/Å | 17.7224(11) |
| c/Å | 17.9592(11) |
| α/° | 65.119(6) |
| β/° | 65.002(5) |
| γ/° | 80.898(5) |
| Volume/Å$^3$ | 4190.3(5) |
| Z | 1 |
| $\rho_{calc}$g/cm$^3$ | 3.003 |
| μ/mm$^{-1}$ | 22.051 |
| F(000) | 3389.0 |
| Crystal size/mm$^3$ | 0.183 × 0.148 × 0.029 |
| Radiation | Mo Kα (λ = 0.71073) |
| 2Θ range for data collection/° | 3.244 to 52.886 |
| Index ranges | $-18 \leq h \leq 20, -22 \leq k \leq 22, -15 \leq l \leq 15$ |
| Reflections collected | 34974 |
| Independent reflections | 4526 [$R_{int}$ = 0.0516, $R_{sigma}$ = 0.0340] |
| Data/restraints/parameters | 4526/366/545 |
| Goodness-of-fit on $F^2$ | 1.055 |
| Final R indexes [I>=2σ (I)] | $R_1$ = 0.0301, $wR_2$ = 0.0584 |
| Final R indexes [all data] | $R_1$ = 0.0506, $wR_2$ = 0.0639 |
| Largest diff. peak/hole / e Å$^{-3}$ | 0.62/-0.54 |

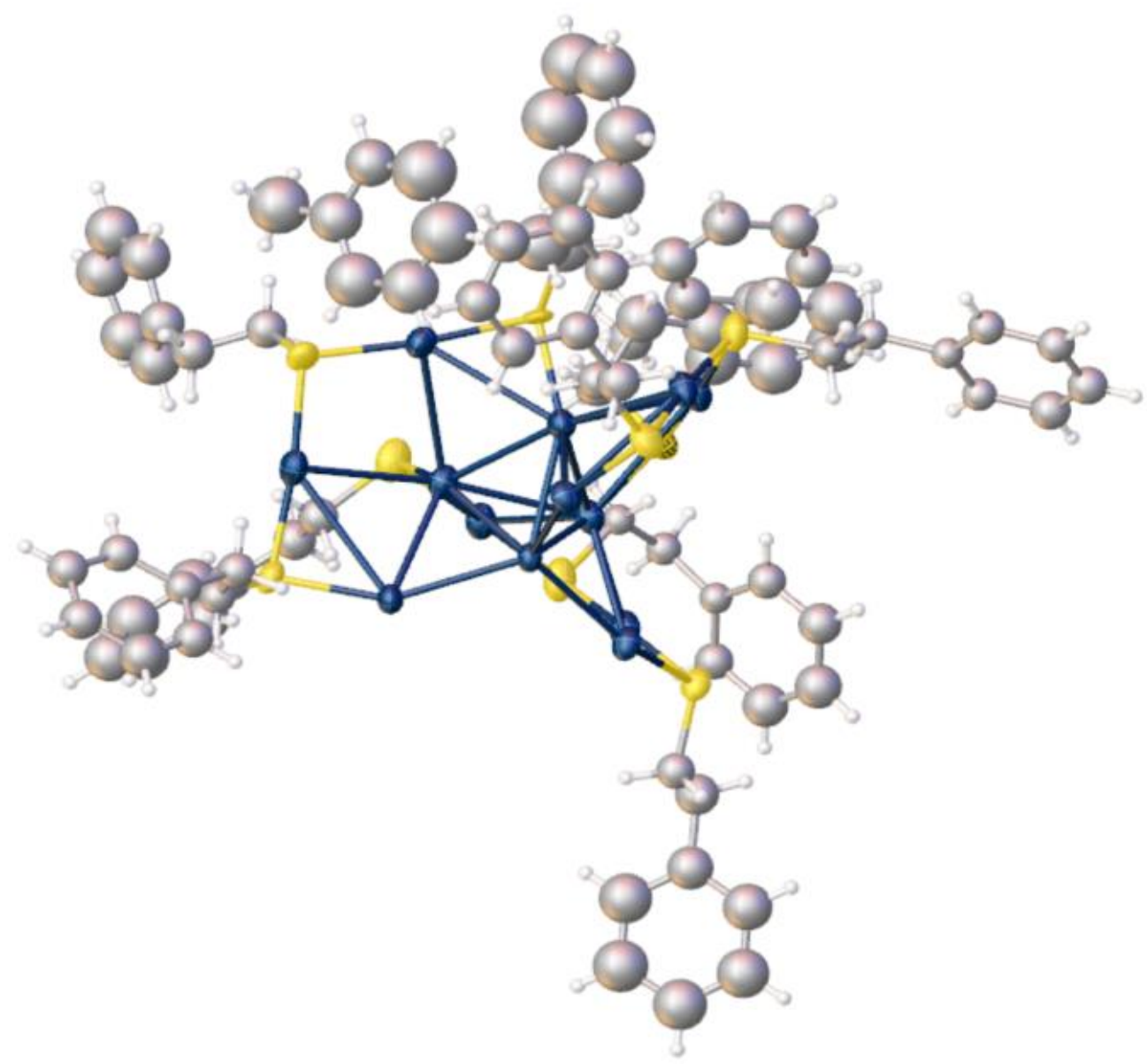

Figure S4. View of the asymmetric unit at 0.4 GPa along $\vec{a}$ with displacement parameters at 50 percent probability level.

- **Checkcif alerts and answers**

_vrf_PLAT029_p01
;
PROBLEM: _diffrn_measured_fraction_theta_full value Low .      0.278 Why?
RESPONSE: These are high-pressure data taken on a home-lab diffractometer. We are limited by the use of the diamond anvil cell and the crystal system is moreover triclinic. Restrained were heavily used for the ligands to try to avoid overfitting.
;
_vrf_PLAT201_p01
;
PROBLEM: Isotropic non-H Atoms in Main Residue(s) Occ>0.5        72 Report
RESPONSE: These are high-pressure data taken on a home-lab diffractometer. We are limited by the use of the diamond anvil cell and the crystal system is moreover triclinic. All C atoms were refined isotropically.
;
_vrf_PLAT342_p01
;
PROBLEM: Low Bond Precision on  C-C Bonds ...............      0.0431 Ang.
RESPONSE: These are high-pressure data taken on a home-lab diffractometer.
;
_vrf_PLAT911_p01
;
PROBLEM: Missing FCF Refl Between Thmin & STh/L=    0.600       9319 Report
RESPONSE: These are high-pressure data taken on a home-lab diffractometer. We are limited by the use of the diamond anvil cell and the crystal system is moreover triclinic.
;
# end Validation Reply Form

## FROM P01 TO P02

Going from P01 to P02, the crystal undergoes a phase transition. The initial disorder (resi 9 and 10) disappears and some ligands (especially resi 5 and 6) change position. This is unfortunately associated with a degradation of the crystal (see image below).

The P01 model refined well except ligands 5 and 6. These were removed from the model, and their positions were reassessed using Fourier difference maps.

S atoms were refined isotropically.

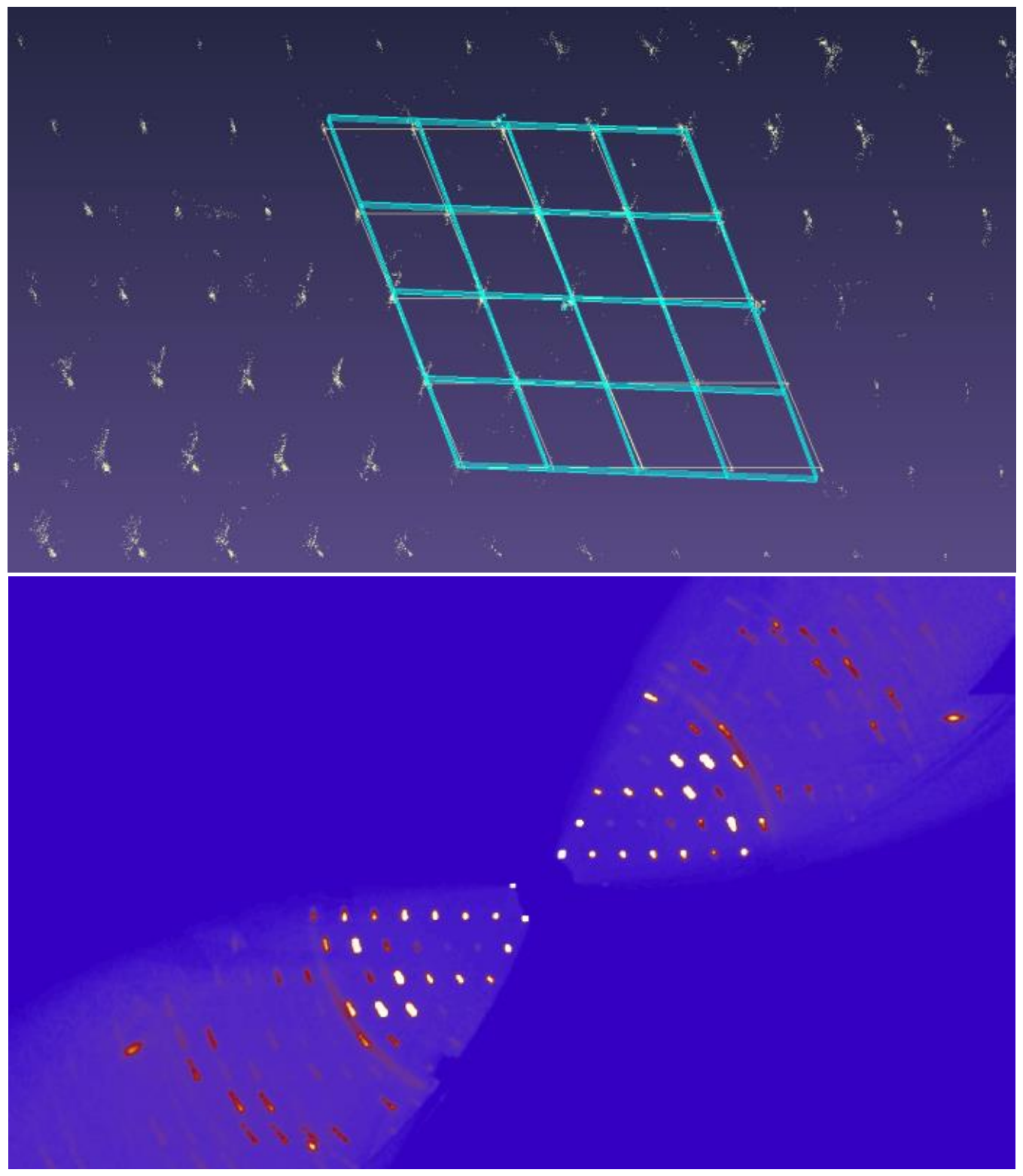

Figure S5. View of the peak hunting window showing the two reciprocal unit cells of the split crystal and view of the reciprocal space reconstruction of the 0kl plane.

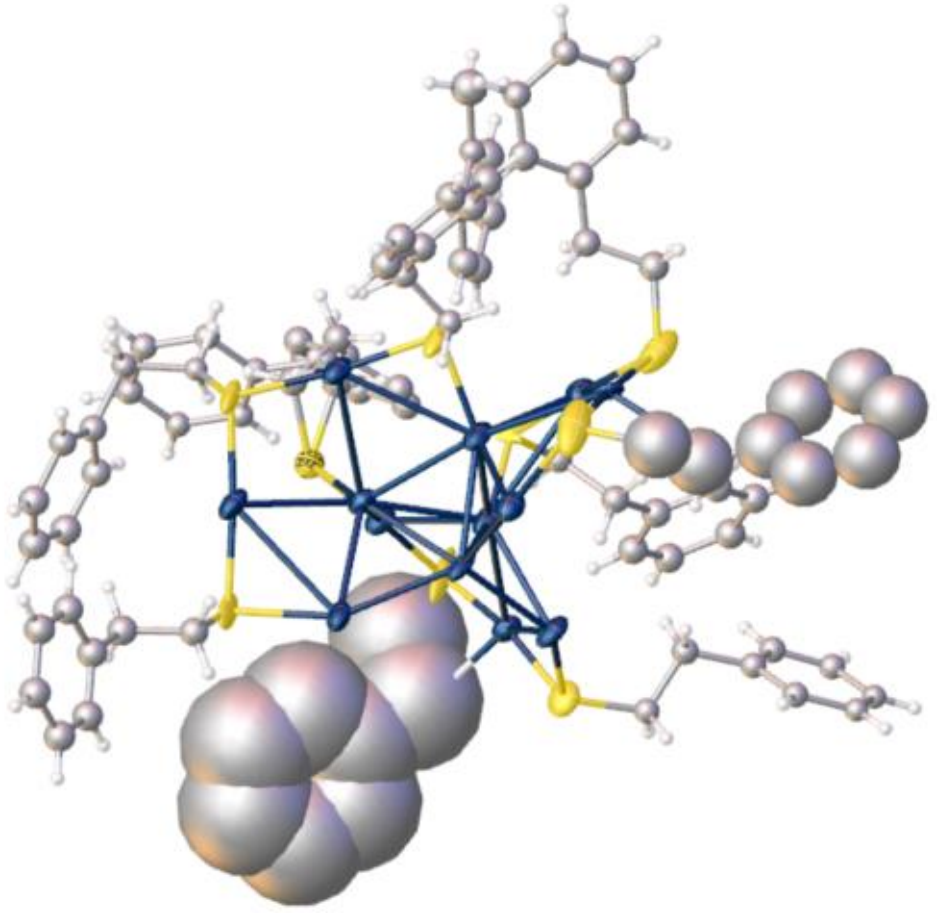

Figure S6. Initial refinement of P02 data from the P01 model. The displacement parameters of ligands 6 and 5 became unusually large, indicated that their positions were too far from the correct locations for the least-squares refinement to converge properly

**P02 FINAL MODEL**

- **Crystal data and structure refinement for P02, 1.2 GPa.**

| | |
|---|---|
| Identification code | P02 |
| Empirical formula | $C_{158}H_{178}Au_{25}S_{18}$ |
| Formula weight | 7578.24 |
| Temperature/K | 298.8(7) |
| Crystal system | triclinic |
| Space group | P-1 |
| a/Å | 16.0761(8) |
| b/Å | 17.1362(15) |
| c/Å | 16.631(3) |
| α/° | 67.369(11) |
| β/° | 65.894(9) |
| γ/° | 83.847(6) |
| Volume/Å$^3$ | 3853.2(8) |
| Z | 1 |
| $\rho_{calc}$g/cm$^3$ | 3.266 |
| μ/mm$^{-1}$ | 23.980 |
| F(000) | 3389.0 |
| Crystal size/mm$^3$ | 0.183 × 0.148 × 0.029 |
| Radiation | Mo Kα (λ = 0.71073) |
| 2Θ range for data collection/° | 3.678 to 57.556 |
| Index ranges | $-16 \leq h \leq 21$, $-22 \leq k \leq 20$, $-13 \leq l \leq 14$ |
| Reflections collected | 7426 |
| Independent reflections | 3277 [$R_{int}$ = 0.0464, $R_{sigma}$ = 0.0672] |
| Data/restraints/parameters | 3277/332/467 |
| Goodness-of-fit on F$^2$ | 1.107 |
| Final R indexes [I>=2σ (I)] | $R_1$ = 0.0650, $wR_2$ = 0.1690 |
| Final R indexes [all data] | $R_1$ = 0.0982, $wR_2$ = 0.1909 |
| Largest diff. peak/hole / e Å$^{-3}$ | 2.04/-1.50 |

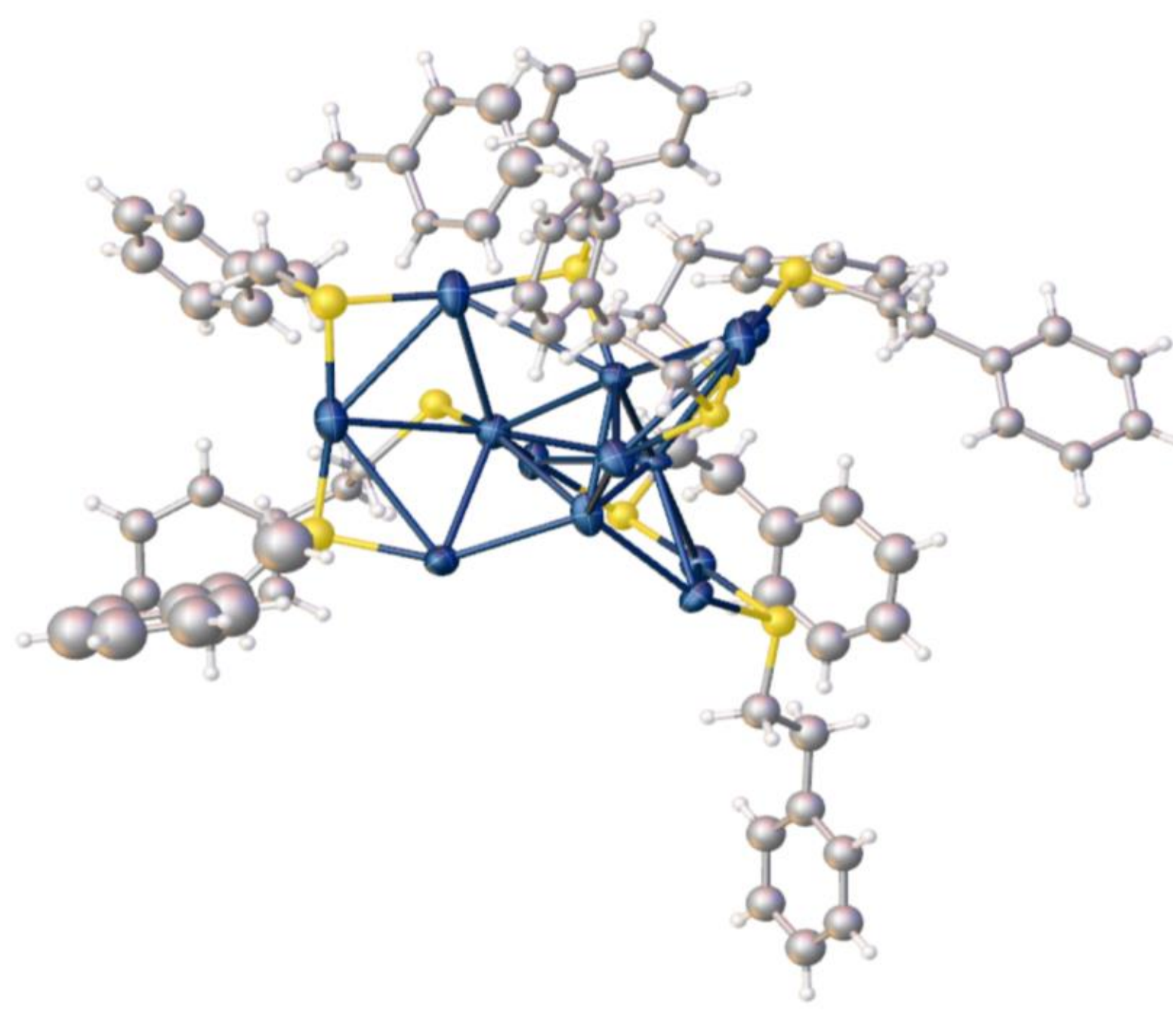

Figure S7. View of the asymmetric unit at 1.2 GPa along $\vec{a}$ with displacement parameters at 50 percent probability level.

- **Checkcif alerts and answers**

# start Validation Reply Form
_vrf_PLAT029_p02
;
PROBLEM: _diffrn_measured_fraction_theta_full value Low . 0.220 Why?
RESPONSE: These are high-pressure data taken on a home-lab diffractometer. We are limited by the use of the diamond anvil cell and the crystal system is moreover triclinic. Restrained were heavily used for the ligands to try to avoid overfitting.
;
_vrf_PLAT088_p02
;
PROBLEM: Poor Data / Parameter Ratio .................... 7.02 Note
RESPONSE: These are high-pressure data taken on a home-lab diffractometer. We are limited by the use of the diamond anvil cell and the crystal system is moreover triclinic. Restrained were heavily used for the ligands to try to avoid overfitting.
;
_vrf_PLAT201_p02
;
PROBLEM: Isotropic non-H Atoms in Main Residue(s) Occ>0.5 81 Report
RESPONSE: These are high-pressure data taken on a home-lab diffractometer. We are limited by the use of the diamond anvil cell and the crystal system is moreover triclinic. Restrained were heavily used for the ligands to try to avoid overfitting. All C and S atoms were refined isotropically.
;
_vrf_PLAT342_p02
;
PROBLEM: Low Bond Precision on C-C Bonds ............... 0.09354 Ang.
RESPONSE:

These are high-pressure data taken on a home-lab diffractometer. We are limited by the use of the diamond anvil cell
;
_vrf_PLAT911_p02
;
PROBLEM: Missing FCF Refl Between Thmin & STh/L= 0.600 9465 Report
RESPONSE:
These are high-pressure data taken on a home-lab diffractometer. We are limited by the use of the diamond anvil cell
;
# end Validation Reply Form

## FROM P02 TO P03

The structure from P02 was used as a starting model. The refinement went smoothly to the final model.

## P03 FINAL MODEL

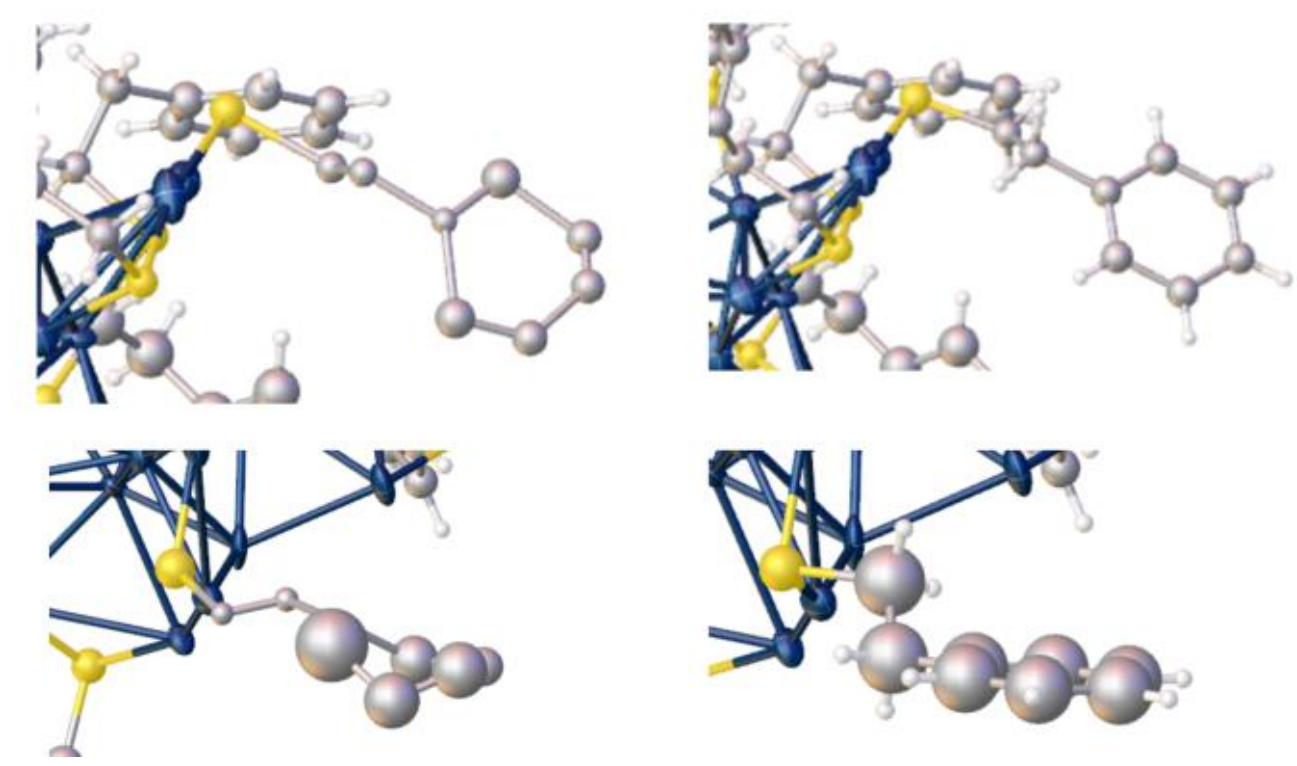

Figure S8. View of the model for residues 5 (top) and 6 (bottom). The initial positions obtained from the Fourier difference map are shown on the left, and the final restrained model is shown on the right.

Alerts indicate short H···H contacts. Although the ligand positions are supported by the data, since they can be relocated from Fourier difference maps after removal (see figure above), the precision remains limited. This is particularly true for residue 5, whose large displacement ellipsoids suggest possible disorder.

- **Crystal data and structure refinement for P03, 2.1 GPa.**

| | |
|---|---|
| Identification code | P03 |
| Empirical formula | $C_{158}H_{178}Au_{25}S_{18}$ |
| Formula weight | 7578.24 |
| Temperature/K | 299.1(3) |
| Crystal system | triclinic |
| Space group | P-1 |
| a/Å | 15.8926(7) |
| b/Å | 16.8498(14) |
| c/Å | 16.212(2) |
| α/° | 67.525(10) |
| β/° | 66.143(8) |
| γ/° | 83.925(5) |
| Volume/Å$^3$ | 3662.7(7) |
| Z | 1 |
| $\rho_{calc}$g/cm$^3$ | 3.436 |
| μ/mm$^{-1}$ | 25.228 |
| F(000) | 3389.0 |
| Crystal size/mm$^3$ | 0.183 × 0.148 × 0.029 |
| Radiation | Mo Kα (λ = 0.71073) |
| 2Θ range for data collection/° | 3.136 to 57.186 |
| Index ranges | $-16 \leq h \leq 20$, $-22 \leq k \leq 22$, $-13 \leq l \leq 13$ |
| Reflections collected | 11466 |
| Independent reflections | 3575 [$R_{int}$ = 0.0445, $R_{sigma}$ = 0.0557] |
| Data/restraints/parameters | 3575/410/467 |
| Goodness-of-fit on $F^2$ | 1.090 |
| Final R indexes [I>=2σ (I)] | $R_1$ = 0.0723, $wR_2$ = 0.1923 |
| Final R indexes [all data] | $R_1$ = 0.1124, $wR_2$ = 0.2377 |
| Largest diff. peak/hole / e $Å^{-3}$ | 2.74/-2.28 |

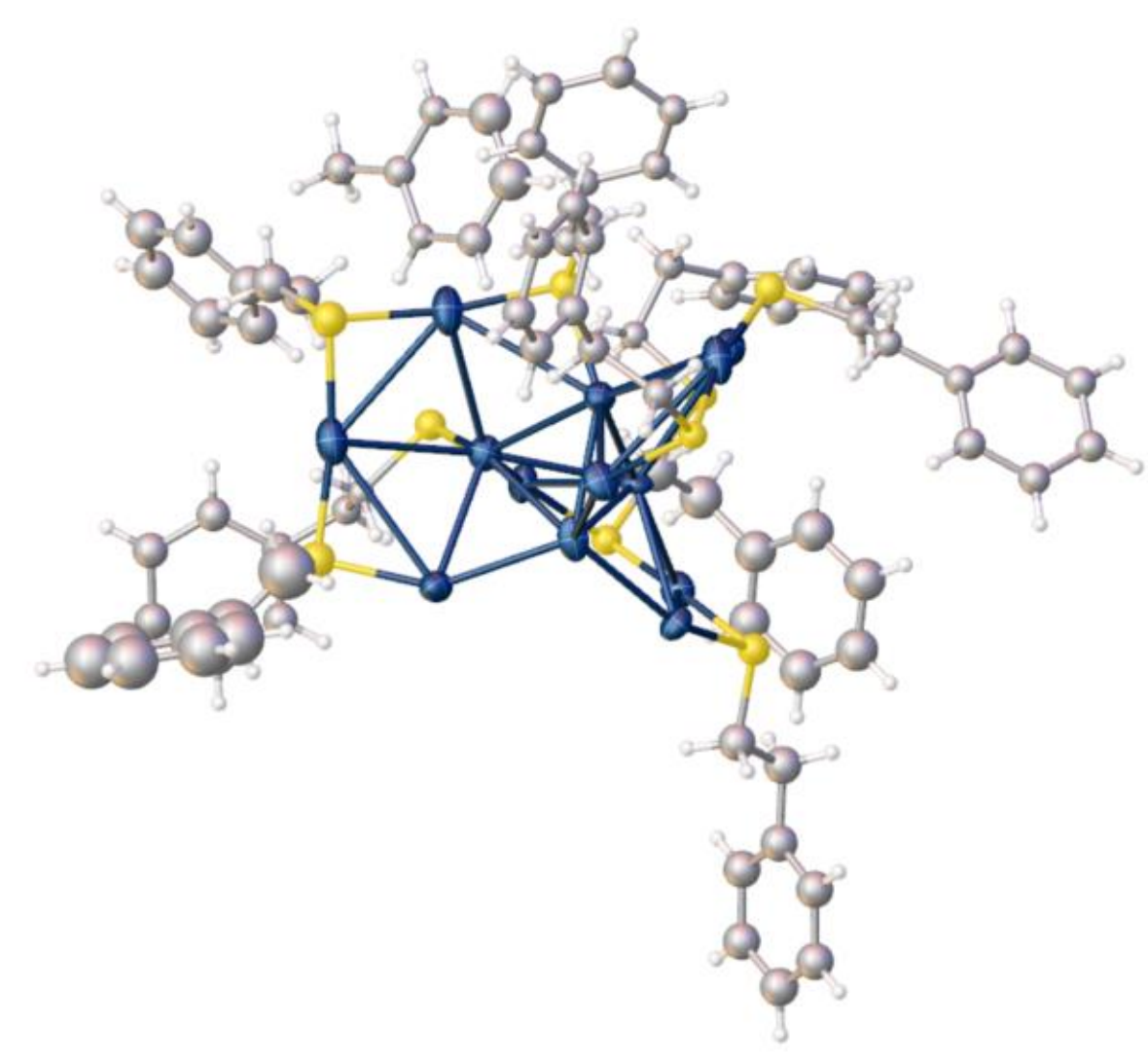

Figure S9. View of the asymmetric unit at 2.1 GPa along $\vec{a}$ with displacement parameters at 50 percent probability level.

- **Checkcif alerts and answers**

# start Validation Reply Form
_vrf_PLAT029_p03
;
PROBLEM: _diffrn_measured_fraction_theta_full value Low . 0.248 Why?
RESPONSE: These are high-pressure data taken on a home-lab diffractometer. We are limited by the use of the diamond anvil cell and the crystal system is moreover triclinic. Restrained were heavily used for the ligands to try to avoid overfitting.
;
_vrf_PLAT342_p03
;
PROBLEM: Low Bond Precision on C-C Bonds ............... 0.10468 Ang.
RESPONSE: These are high-pressure data taken on a home-lab diffractometer. We are limited by the use of the diamond anvil cell
;
_vrf_PLAT410_p03
;

PROBLEM: Short Intra H...H Contact H20A_6 ..H21A_5 . 1.33 Ang.
RESPONSE:
Although the ligand positions are supported by the data, since they can be relocated from Fourier difference maps after removal, the precision remains limited. This is particularly true for residue 5, whose large displacement ellipsoids suggest possible disorder.
;
_vrf_PLAT088_p03
;
PROBLEM: Poor Data / Parameter Ratio .................... 7.66 Note

RESPONSE: These are high-pressure data taken on a home-lab diffractometer. We are limited by the use of the diamond anvil cell and the crystal system is moreover triclinic. Restrained were heavily used for the ligands to try to avoid overfitting.
;
_vrf_PLAT201_p03
;
PROBLEM: Isotropic non-H Atoms in Main Residue(s) Occ>0.5 81 Report
RESPONSE: These are high-pressure data taken on a home-lab diffractometer. We are limited by the use of the diamond anvil cell and the crystal system is moreover triclinic. Restrained were heavily used for the ligands to try to avoid overfitting. All C and S atoms were refined
;
_vrf_PLAT911_p03
;
PROBLEM: Missing FCF Refl Between Thmin & STh/L= 0.600 8287 Report
These are high-pressure data taken on a home-lab diffractometer. We are limited by the use of the diamond anvil cell
# end Validation Reply Form

## FROM P03 TO P04

The data quality continues to slowly deteriorate, due to the splitting of the crystals. However, the model from P03 could be used as a starting model.

## P04 FINAL MODEL

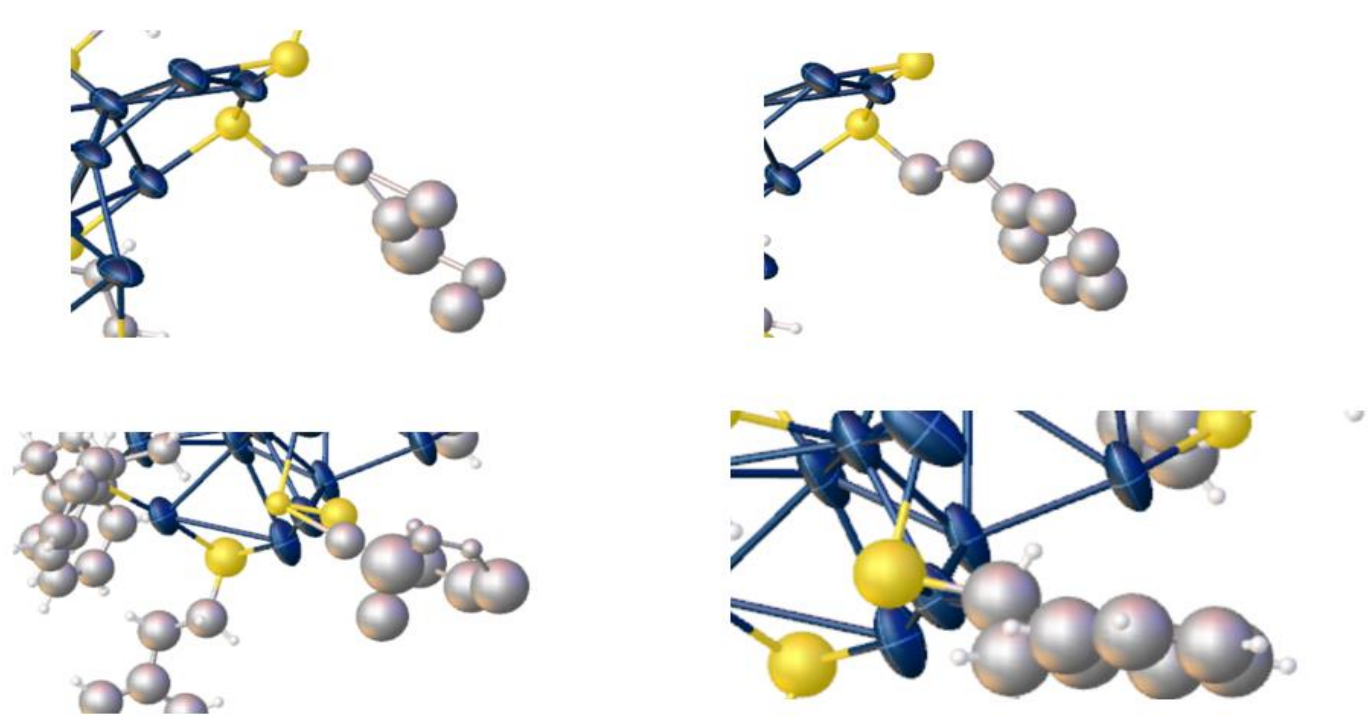

Figure S10. View of the model for residues 6 (top) and 5 (bottom). The initial positions obtained from the Fourier difference map are shown on the left, and the final restrained model is shown on the right. For resiudue 5, 2 position could be located for the S atom.

Alerts indicate short H···H contacts. Although the ligand positions are supported by the data, since they can be relocated from Fourier difference maps after removal (see figure above), the precision remains limited. This is particularly true for residue 5, whose large displacement ellipsoids suggest possible disorder. When trying to relocate residue 5, an alternative position for S starts appearing. However, no clear density was seen for the rest of the alternative position of the ligand so that only one ligand position was kept in the final model.

- **Crystal data and structure refinement for P04, 3.1 GPa.**

| | |
|---|---|
| Identification code | P04 |
| Empirical formula | $C_{158}H_{178}Au_{25}S_{18}$ |
| Formula weight | 7578.24 |
| Temperature/K | 299.0(4) |
| Crystal system | triclinic |
| Space group | P-1 |
| a/Å | 15.5717(10) |
| b/Å | 16.7182(17) |
| c/Å | 15.874(2) |
| α/° | 67.550(11) |
| β/° | 67.172(9) |
| γ/° | 83.854(7) |
| Volume/Å$^3$ | 3516.1(7) |
| Z | 1 |
| $\rho_{calc}$g/cm$^3$ | 3.579 |
| μ/mm$^{-1}$ | 26.280 |
| F(000) | 3389.0 |
| Crystal size/mm$^3$ | 0.183 × 0.148 × 0.029 |
| Radiation | Mo Kα (λ = 0.71073) |
| 2Θ range for data collection/° | 3.174 to 58.172 |
| Index ranges | $-18 \leq h \leq 20$, $-22 \leq k \leq 22$, $-12 \leq l \leq 12$ |
| Reflections collected | 23759 |
| Independent reflections | 3519 [$R_{int}$ = 0.0913, $R_{sigma}$ = 0.0558] |
| Data/restraints/parameters | 3519/410/467 |
| Goodness-of-fit on $F^2$ | 1.035 |
| Final R indexes [I>=2σ (I)] | $R_1$ = 0.0713, $wR_2$ = 0.1924 |
| Final R indexes [all data] | $R_1$ = 0.1190, $wR_2$ = 0.2271 |
| Largest diff. peak/hole / e Å$^{-3}$ | 1.87/-2.29 |

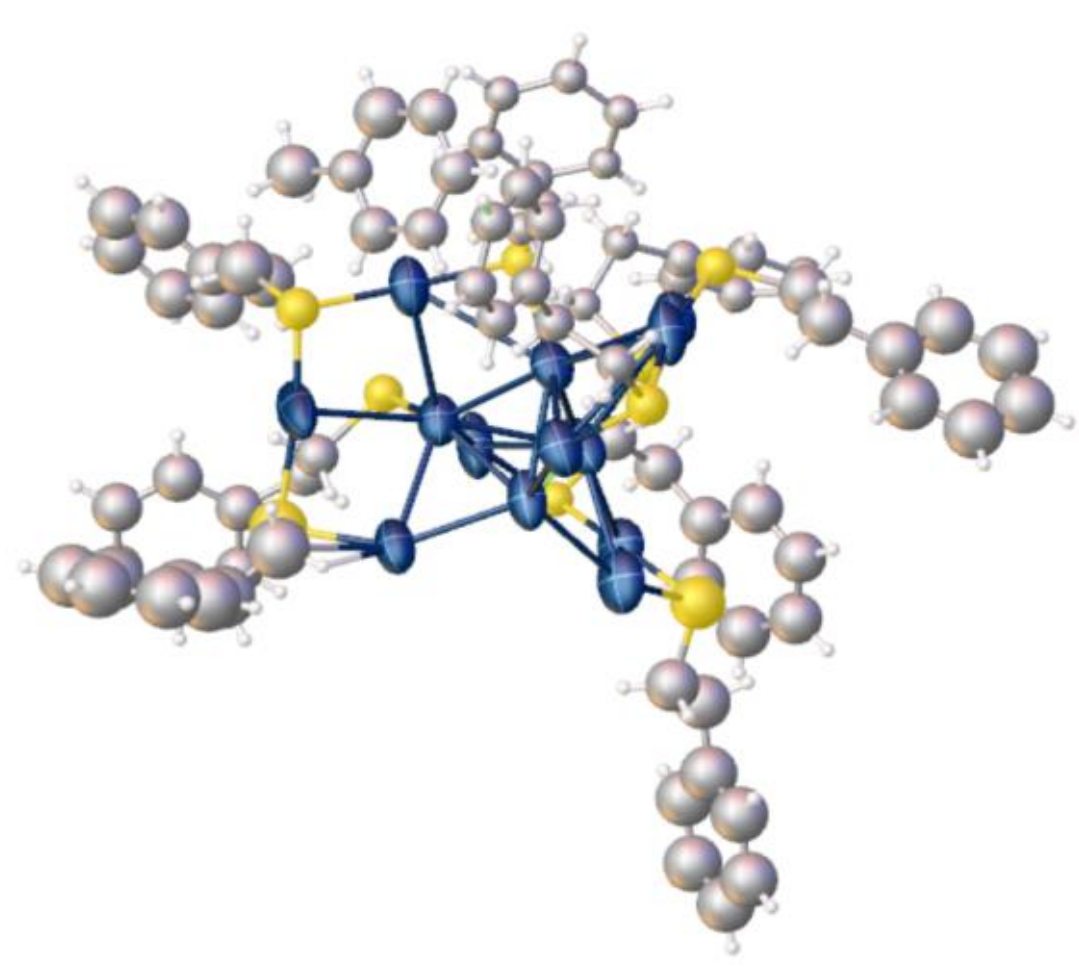

Figure S11. View of the asymmetric unit at 3.1 GPa along $\vec{a}$ with displacement parameters at 50 percent probability level.

- **Checkcif alerts and answers**

# start Validation Reply Form

_vrf_PLAT029_p04

;

PROBLEM: _diffrn_measured_fraction_theta_full value Low . 0.239 Why?

RESPONSE: These are high-pressure data taken on a home-lab diffractometer. We are limited by the use of the diamond anvil cell and the crystal system is moreover triclinic. Restrained were heavily used for the ligands to try to avoid overfitting.

;

_vrf_PLAT342_p04

;

PROBLEM: Low Bond Precision on C-C Bonds ............... 0.10089 Ang.

RESPONSE: These are high-pressure data taken on a home-lab diffractometer. We are limited by the use of the diamond anvil cell

;
_vrf_PLAT410_p04
;
PROBLEM: Short Intra H...H Contact H20A_6 ..H21A_5 . 1.28 Ang.
RESPONSE: Although the ligand positions are supported by the data, since they can be relocated from Fourier difference maps after removal, the precision remains limited. This is particularly true for residue 5, where a disorder may be present.
;
_vrf_PLAT411_p04
;
PROBLEM: Short Inter H...H Contact H23_3 ..H25_5 . 1.50 Ang.
RESPONSE: Although the ligand positions are supported by the data, since they can be relocated from Fourier difference maps after removal, the precision remains limited. This is particularly true for residue 5, where a disorder may be present.
;
_vrf_PLAT088_p04
;
PROBLEM: Poor Data / Parameter Ratio .................... 7.54 Note
RESPONSE: These are high-pressure data taken on a home-lab diffractometer. We are limited by the use of the diamond anvil cell and the crystal system is moreover triclinic. Restrained were heavily used for the ligands to try to avoid overfitting.
;
_vrf_PLAT201_p04
;
PROBLEM: Isotropic non-H Atoms in Main Residue(s) Occ>0.5 81 Report
RESPONSE: These are high-pressure data taken on a home-lab diffractometer. We are limited by the use of the diamond anvil cell and the crystal system is moreover triclinic. Restrained were heavily used for the ligands to try to avoid overfitting. All C and S atoms were refined isotropically.
;
_vrf_PLAT911_p04
;
PROBLEM: Missing FCF Refl Between Thmin & STh/L= 0.600 7838 Report
RESPONSE: These are high-pressure data taken on a home-lab diffractometer. We are limited by the use of the diamond anvil cell
;
# end Validation Reply Form

**FROM P04 TO P05**

The data quality continues to slowly deteriorate, due to the splitting of the crystals. However, the model from P03 could be used as a starting model. Residue 5 starts to be difficult to localize in the Fourier difference maps, so that its position is getting more uncertain.

**P04 FINAL MODEL**

- **Crystal data and structure refinement for P05, 3.6 GPa.**

| | |
|---|---|
| Identification code | P05 |
| Empirical formula | $C_{158}H_{178}Au_{25}S_{18}$ |
| Formula weight | 7578.24 |
| Temperature/K | 299.1(3) |
| Crystal system | triclinic |
| Space group | P-1 |
| a/Å | 15.4795(10) |
| b/Å | 16.6235(18) |
| c/Å | 15.747(2) |
| α/° | 67.694(12) |
| β/° | 67.364(10) |
| γ/° | 83.943(7) |
| Volume/Å$^3$ | 3456.2(8) |
| Z | 1 |
| $\rho_{calc}$g/cm$^3$ | 3.641 |
| μ/mm$^{-1}$ | 26.735 |
| F(000) | 3389.0 |
| Crystal size/mm$^3$ | 0.183 × 0.148 × 0.029 |
| Radiation | Mo Kα (λ = 0.71073) |
| 2Θ range for data collection/° | 3.196 to 57.304 |
| Index ranges | $-12 \leq h \leq 20$, $-21 \leq k \leq 22$, $-12 \leq l \leq 12$ |
| Reflections collected | 12175 |
| Independent reflections | 3244 [$R_{int}$ = 0.0542, $R_{sigma}$ = 0.0574] |
| Data/restraints/parameters | 3244/410/468 |
| Goodness-of-fit on F$^2$ | 1.063 |
| Final R indexes [I>=2σ (I)] | $R_1$ = 0.0853, $wR_2$ = 0.2292 |
| Final R indexes [all data] | $R_1$ = 0.1334, $wR_2$ = 0.2676 |
| Largest diff. peak/hole / e Å$^{-3}$ | 1.72/-1.85 |

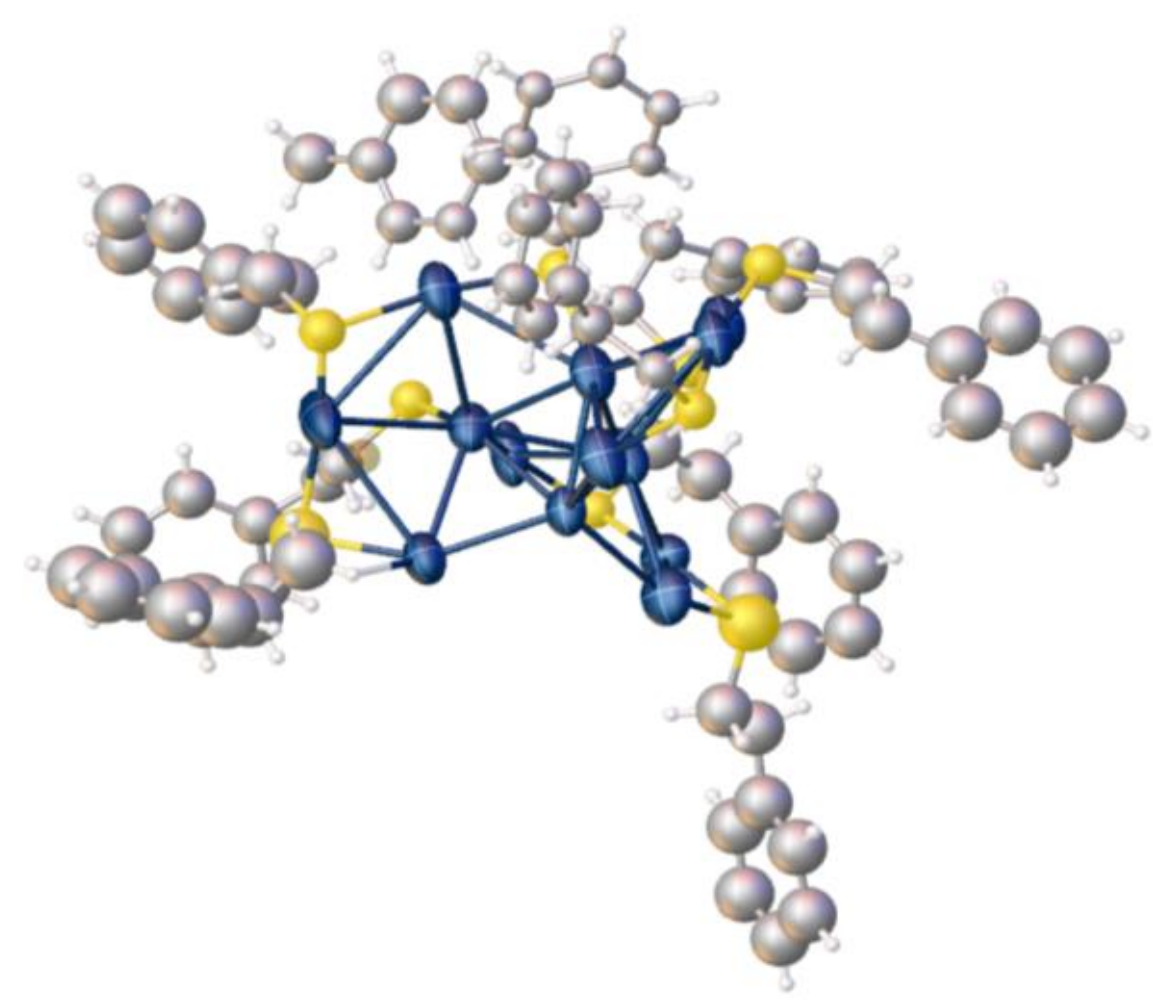

Figure S12. View of the asymmetric unit at 3.6 GPa along $\vec{a}$ with displacement parameters at 50 percent probability level.

- **Checkcif alerts and answers**

# start Validation Reply Form
_vrf_PLAT029_p05
;
PROBLEM: _diffrn_measured_fraction_theta_full value Low . 0.237 Why?
RESPONSE: These are high-pressure data taken on a home-lab diffractometer. We are limited by the use of the diamond anvil cell and the crystal system is moreover triclinic. Restrained were heavily used for the ligands to try to avoid overfitting.
;
_vrf_PLAT342_p05
;
PROBLEM: Low Bond Precision on C-C Bonds ............... 0.12329 Ang.
RESPONSE: These are high-pressure data taken on a home-lab diffractometer. We are limited by the use of the diamond anvil cell.
;
_vrf_PLAT410_p05
;
PROBLEM: Short Intra H...H Contact H20A_6 ..H21A_5 . 1.44 Ang.
RESPONSE : Although the ligand positions are partly supported by the data, since they can m be somehow relocated from Fourier difference maps after removal, the precision remains limited and gets worth and worth as the pressure increases. This is particularly true for residue 5, where a disorder may be present.
;
_vrf_PLAT411_p05
;
PROBLEM: Short Inter H...H Contact H26_1 ..H26_6 . 1.78 Ang.

RESPONSE : Although the ligand positions are partly supported by the data, since they can m be somehow relocated from Fourier difference maps after removal,the precision remains limited and gets worth and worth as the pressure increases. This is particularly true for residue 5, where a disorder may be present.

;
_vrf_PLAT413_p05
;
PROBLEM: Short Inter XH3 .. XHn H2C ..H24_2 . 1.85 Ang.

RESPONSE : Although the ligand positions are partly supported by the data, since they can m be somehow relocated from Fourier difference maps after removal, the precision remains limited and gets worth and worth as the pressure increases. This is particularly true for residue 5, where a disorder may be present.
;
_vrf_PLAT088_p05
;
PROBLEM: Poor Data / Parameter Ratio .................... 6.93 Note

RESPONSE: These are high-pressure data taken on a home-lab diffractometer. We are limited by the use of the diamond anvil cell and the crystal system is moreover triclinic. Restrained were heavily used for the ligands to try to avoid overfitting.
;
_vrf_PLAT201_p05
;
PROBLEM: Isotropic non-H Atoms in Main Residue(s) Occ>0.5 81 Report
RESPONSE: These are high-pressure data taken on a home-lab diffractometer. We are limited by the use of the diamond anvil cell and the crystal system is moreover triclinic. Restrained were heavily used for the ligands to try to avoid overfitting. All C and S atoms were refined isotropically.
;
_vrf_PLAT911_p05
;
PROBLEM: Missing FCF Refl Between Thmin & STh/L= 0.600 7630 Report
RESPONSE: These are high-pressure data taken on a home-lab diffractometer. We are limited by the use of the diamond anvil cell
;
# end Validation Reply Form

## FROM P05 TO P06

The data quality continues to deteriorate, and the ligand positions exhibit increasing disorder. In the structural model, three ligands were modelled as disordered over two sites: ligand 5 (refined over positions lig5 and lig15 with occupancies of 0.5), ligand 2 (lig2 and lig12, with occupancies of 0.5), and ligand 3 (lig3 and lig13, with occupancies of 0.5).

An ISOR restraint was applied to the displacement parameters of the central Au atom. The refinement of the atomic displacement parameters is highly sensitive to minor variations in the integration procedure, particularly to changes in the integration box size.

## P06 FINAL MODEL

- **Crystal data and structure refinement for P06, 4.9 GPa.**

| | |
|---|---|
| Identification code | P06 |
| Empirical formula | $C_{158}H_{178}Au_{25}S_{18}$ |
| Formula weight | 7578.24 |
| Temperature/K | 299.1(5) |
| Crystal system | triclinic |
| Space group | P-1 |
| a/Å | 15.1625(11) |
| b/Å | 16.3374(16) |
| c/Å | 15.472(4) |
| α/° | 68.046(15) |
| β/° | 68.192(13) |
| γ/° | 84.492(7) |
| Volume/Å$^3$ | 3296.3(10) |
| Z | 1 |
| $\rho_{calc}$g/cm$^3$ | 3.818 |
| μ/mm$^{-1}$ | 28.032 |
| F(000) | 3389.0 |
| Crystal size/mm$^3$ | 0.183 × 0.148 × 0.029 |
| Radiation | Mo Kα (λ = 0.71073) |
| 2Θ range for data collection/° | 2.896 to 56.394 |
| Index ranges | $-15 \leq h \leq 19$, $-21 \leq k \leq 21$, $-9 \leq l \leq 9$ |
| Reflections collected | 16340 |
| Independent reflections | 2746 [$R_{int}$ = 0.0619, $R_{sigma}$ = 0.0431] |
| Data/restraints/parameters | 2746/554/576 |
| Goodness-of-fit on F$^2$ | 1.045 |
| Final R indexes [I>=2σ (I)] | $R_1$ = 0.0739, w$R_2$ = 0.1973 |
| Final R indexes [all data] | $R_1$ = 0.1115, w$R_2$ = 0.2296 |
| Largest diff. peak/hole / e Å$^{-3}$ | 1.46/-1.50 |

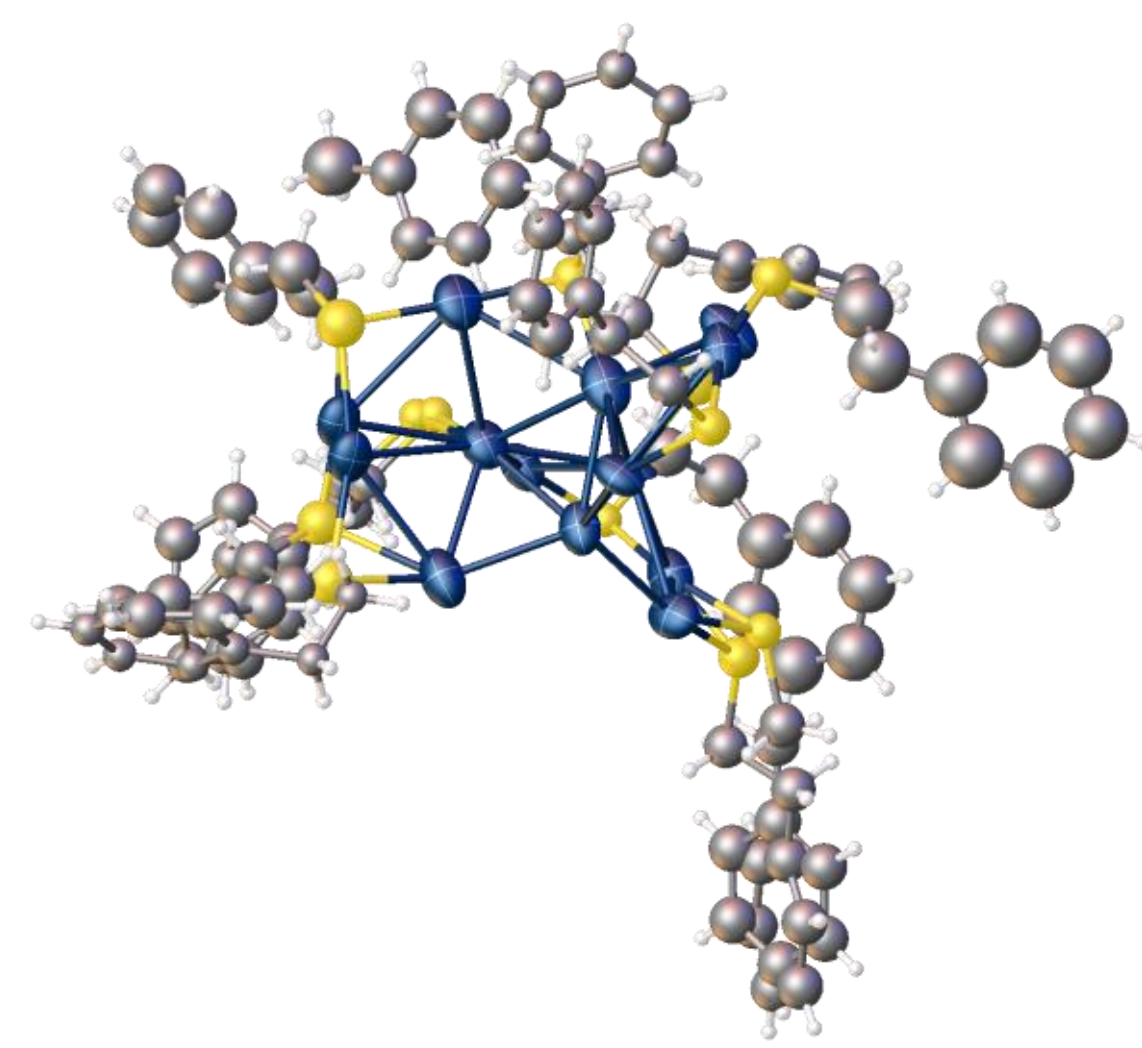

Figure S13. View of the asymmetric unit at 4.9 GPa along $\vec{a}$ with displacement parameters at 50 percent probability level.

- **Checkcif alerts and answers**

_vrf_PLAT029_p06
;
PROBLEM: _diffrn_measured_fraction_theta_full value Low . 0.211 Why?
RESPONSE: These are high-pressure data taken on a home-lab diffractometer. We are limited by the use of the diamond anvil cell and the crystal system is moreover triclinic. Restrained were heavily used for the ligands to try to avoid overfitting.
;
_vrf_PLAT088_p06
;
PROBLEM: Poor Data / Parameter Ratio .................... 4.77 Note
RESPONSE: These are high-pressure data taken on a home-lab diffractometer. We are limited by the use of the diamond anvil cell and the crystal system is moreover triclinic. Restrained were heavily used for the ligands to try to avoid overfitting.
;
_vrf_PLAT342_p06
;
PROBLEM: Low Bond Precision on C-C Bonds ............... 0.11509 Ang.
RESPONSE: These are high-pressure data taken on a home-lab diffractometer. We are limited by the use of the diamond anvil cell.
;
_vrf_PLAT201_p06
;
PROBLEM: Isotropic non-H Atoms in Main Residue(s) Occ>0.5 54 Report
RESPONSE: These are high-pressure data taken on a home-lab diffractometer. We are limited by the use of the diamond anvil cell and the crystal system is moreover triclinic. Restrained were heavily used for the ligands to try to avoid overfitting. All C and S atoms were refined isotropically.
;
_vrf_PLAT411_p06
;

PROBLEM: Short Inter H...H Contact  H21B_1   ..H20A_7   .       1.96 Ang.
RESPONSE : Although the ligand positions are partly supported by the data, since they can be somehow relocated  from Fourier difference maps after removal, the precision remains limited and gets worth and worth as the pressure increases.
;
_vrf_PLAT413_p06
;
PROBLEM: Short Inter XH3 .. XHn     H2A      ..H24_6    .       1.94 Ang.
RESPONSE : Although the ligand positions are partly supported by the data, since they can be somehow relocated  from Fourier difference maps after removal, the precision remains limited and gets worth and worth as the pressure increases.
;
_vrf_PLAT911_p06
;
PROBLEM: Missing FCF Refl Between Thmin & STh/L=    0.600       5839 Report
RESPONSE: These are high-pressure data taken on a home-lab diffractometer. We are limited by the use of the diamond anvil cell

## FROM P06 TO P07

The data quality continues to deteriorate. The same model was kept, although possible disorder affecting ligands 1 and 7 is suggested by their large displacement parameters 1 and the short H···H contact (also involving ligand 7 but the latter exhibiting smaller displacement parameters, significant disorder is less likely)

## P07 FINAL MODEL

- **Crystal data and structure refinement for P07, 6.7 GPa.**

| | |
|---|---|
| Identification code | P07 |
| Empirical formula | $C_{158}H_{178}Au_{25}S_{18}$ |
| Formula weight | 7578.24 |
| Temperature/K | 299.3(5) |
| Crystal system | triclinic |
| Space group | P-1 |
| a/Å | 14.9829(11) |
| b/Å | 16.1133(15) |
| c/Å | 15.262(3) |
| α/° | 68.270(14) |
| β/° | 68.274(13) |
| γ/° | 84.630(7) |
| Volume/Å$^3$ | 3175.6(8) |
| Z | 1 |
| $\rho_{calc}$g/cm$^3$ | 3.963 |
| μ/mm$^{-1}$ | 29.097 |
| F(000) | 3389.0 |
| Crystal size/mm$^3$ | 0.183 × 0.148 × 0.029 |
| Radiation | Mo Kα (λ = 0.71073) |
| 2Θ range for data collection/° | 2.93 to 58.66 |
| Index ranges | $-15 \le h \le 19$, $-21 \le k \le 21$, $-9 \le l \le 9$ |
| Reflections collected | 19751 |
| Independent reflections | 2699 [$R_{int}$ = 0.0971, $R_{sigma}$ = 0.0508] |
| Data/restraints/parameters | 2699/555/576 |
| Goodness-of-fit on F$^2$ | 1.134 |
| Final R indexes [I>=2σ (I)] | $R_1$ = 0.0825, w$R_2$ = 0.2190 |
| Final R indexes [all data] | $R_1$ = 0.1141, w$R_2$ = 0.2475 |
| Largest diff. peak/hole / e Å$^{-3}$ | 1.79/-1.32 |

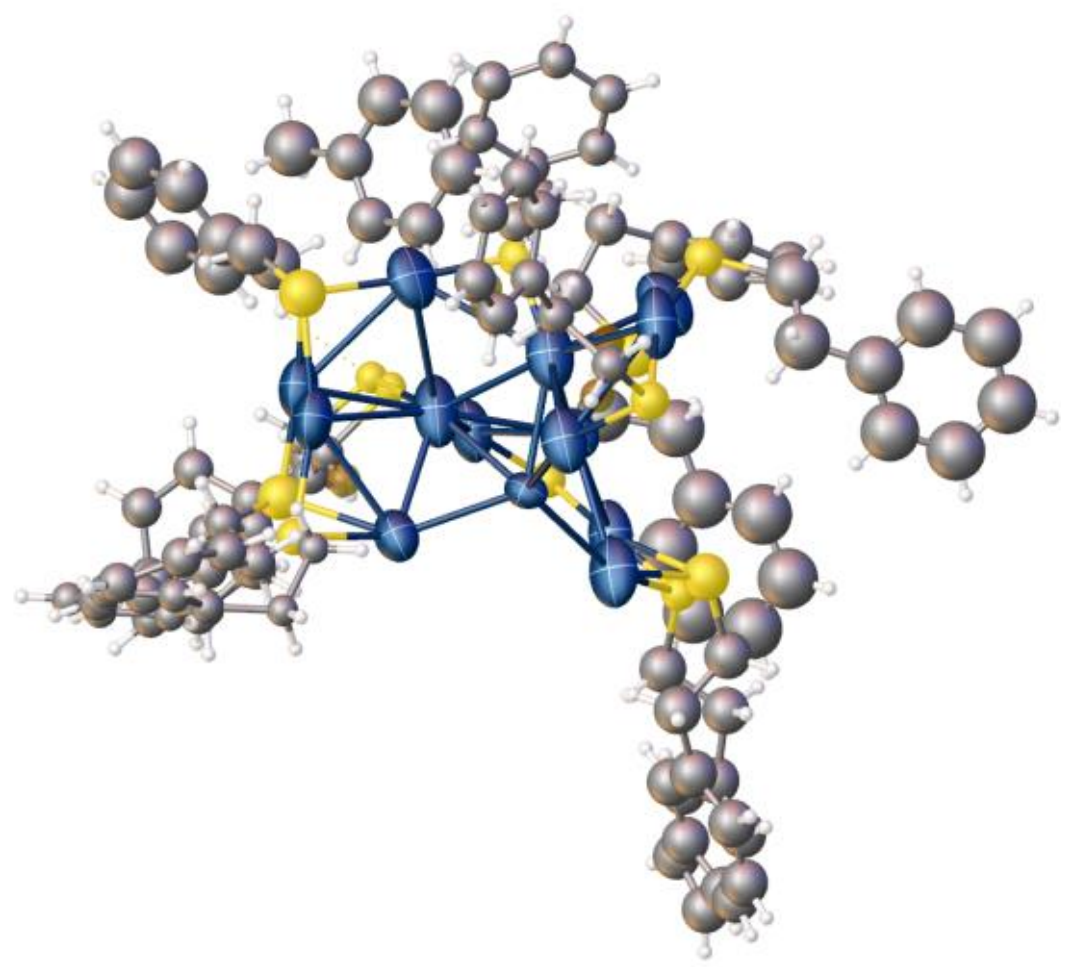

Figure S14. View of the asymmetric unit at 6.7 GPa along $\vec{a}$ with displacement parameters at 50 percent probability level.

- **Checkcif alerts and answers**

# start Validation Reply Form
_vrf_PLAT029_p07
;
PROBLEM: _diffrn_measured_fraction_theta_full value Low .    0.214 Why?
RESPONSE: These are high-pressure data taken on a home-lab diffractometer. We are limited by the use of the diamond anvil cell and the crystal system is moreover triclinic. Restrained were heavily used for the ligands to try to avoid overfitting.
;
_vrf_PLAT088_p07
;
PROBLEM: Poor Data / Parameter Ratio ....................    4.69 Note
RESPONSE: These are high-pressure data taken on a home-lab diffractometer. We are limited by the use of the diamond anvil cell and the crystal system is moreover triclinic. Restrained were heavily used for the ligands to try to avoid overfitting.
;
_vrf_PLAT342_p07
;
PROBLEM: Low Bond Precision on C-C Bonds ...............    0.11436 Ang.
RESPONSE: These are high-pressure data taken on a home-lab diffractometer. We are limited by the use of the diamond anvil cell.
;
;
_vrf_PLAT411_p07
;
PROBLEM: Short Inter H...H Contact H20A_1 ..H20A_7 .    1.77 Ang.
RESPONSE : Although the ligand positions are partly supported by the data, since they can be somehow relocated from Fourier difference maps after removal, the precision remains limited and gets worth and worth as the pressure increases, Lig 1 and 6 may be disordered , as shown by their large displacement ellipsoids. However, due to the limited completeness on the data, we did not introduce a disordered model for them
_vrf_PLAT201_p07

;
PROBLEM: Isotropic non-H Atoms in Main Residue(s) Occ>0.5 54 Report
RESPONSE: These are high-pressure data taken on a home-lab diffractometer. We are limited by the use of the diamond anvil cell and the crystal system is moreover triclinic. Restrained were heavily used for the ligands to try to avoid overfitting. All C and S atoms were refined isotropically.
;
_vrf_PLAT241_p07
;
PROBLEM: High MainResAtom Ueq as Compared to Neighbours Au3 Check
This atom may be slightly disordered; however, more generally, as the data quality deteriorates with increasing pressure, the displacement ellipsoids of the metal atoms become less regular. They are also highly sensitive to the integration parameters and to the applied absorption correction.
_vrf_PLAT242_p07
;
PROBLEM: Low MainResAtom Ueq as Compared to Neighbours Au4 Check
RESPONSE: as the data quality deteriorates with increasing pressure, the displacement ellipsoids of the metal atoms become less regular. They are also highly sensitive to the integration parameters and to the applied absorption correction.
;
_vrf_PLAT250_p07
;
PROBLEM: Large U3/U1 Ratio for <U(i,j)> Tensor(Resd 1) 4.4 Note
RESPONSE: as the data quality deteriorates with increasing pressure, the displacement ellipsoids of the metal atoms become less regular. They are also highly sensitive to the integration parameters and to the applied absorption correction.
;
_vrf_PLAT410_p07
;
PROBLEM: Short Intra H...H Contact H24_1 ..H24_7 . 1.82 Ang.
RESPONSE : Although the ligand positions are partly supported by the data, since they can be somehow relocated from Fourier difference maps after removal, the precision remains limited and gets worth and worth as the pressure increases, Lig 1 and 6 may be disordered , as shown by their large displacement ellipsoids. However, due to the limited completeness on the data, we did not introduce a disordered model for them
;
_vrf_PLAT911_p07
;
PROBLEM: Missing FCF Refl Between Thmin & STh/L= 0.600 5671 Report
RESPONSE : Although the ligand positions are partly supported by the data, since they can be somehow relocated from Fourier difference maps after removal, the precision remains limited and gets worth and worth as the pressure increases, Lig 1 and 6 may be disordered , as shown by their large displacement ellipsoids. However, due to the limited completeness on the data, we did not introduce a disordered model for them
;
# end Validation Reply Form

**FROM P07 TO P08**

The model from P07 was used for P08. Ligands 6 and 1 show very large displacement parameters. They were removed from the model and repositioned using difference Fourier maps (with some imagination), but this did not significantly modify the model nor improve the displacement parameters. Disorder is likely present around these ligands, as also suggested by the unusually short H···H contact; however, the data quality does not allow a reliable or conclusive disorder model.

**P08 FINAL MODEL**

- **Crystal data and structure refinement for P08, 8.4 GPa.**

| | |
|---|---|
| Identification code | P08 |
| Empirical formula | $C_{158}H_{178}Au_{25}S_{18}$ |
| Formula weight | 7578.24 |
| Temperature/K | 299.2(4) |
| Crystal system | triclinic |
| Space group | P-1 |
| a/Å | 14.8674(8) |
| b/Å | 15.9644(11) |
| c/Å | 15.126(3) |
| α/° | 68.439(11) |
| β/° | 68.423(10) |
| γ/° | 84.740(5) |
| Volume/Å$^3$ | 3100.9(7) |
| Z | 1 |
| $\rho_{calc}$g/cm$^3$ | 4.058 |
| μ/mm$^{-1}$ | 29.798 |
| F(000) | 3389.0 |
| Crystal size/mm$^3$ | 0.183 × 0.148 × 0.029 |
| Radiation | Mo Kα (λ = 0.71073) |
| 2Θ range for data collection/° | 2.95 to 55.056 |
| Index ranges | $-12 \leq h \leq 18$, $-20 \leq k \leq 20$, $-7 \leq l \leq 7$ |
| Reflections collected | 13241 |
| Independent reflections | 2296 [$R_{int}$ = 0.0594, $R_{sigma}$ = 0.0438] |
| Data/restraints/parameters | 2296/557/576 |
| Goodness-of-fit on F$^2$ | 1.057 |
| Final R indexes [I>=2σ (I)] | $R_1$ = 0.0695, $wR_2$ = 0.1829 |
| Final R indexes [all data] | $R_1$ = 0.1034, $wR_2$ = 0.2100 |
| Largest diff. peak/hole / e Å$^{-3}$ | 1.68/-1.18 |

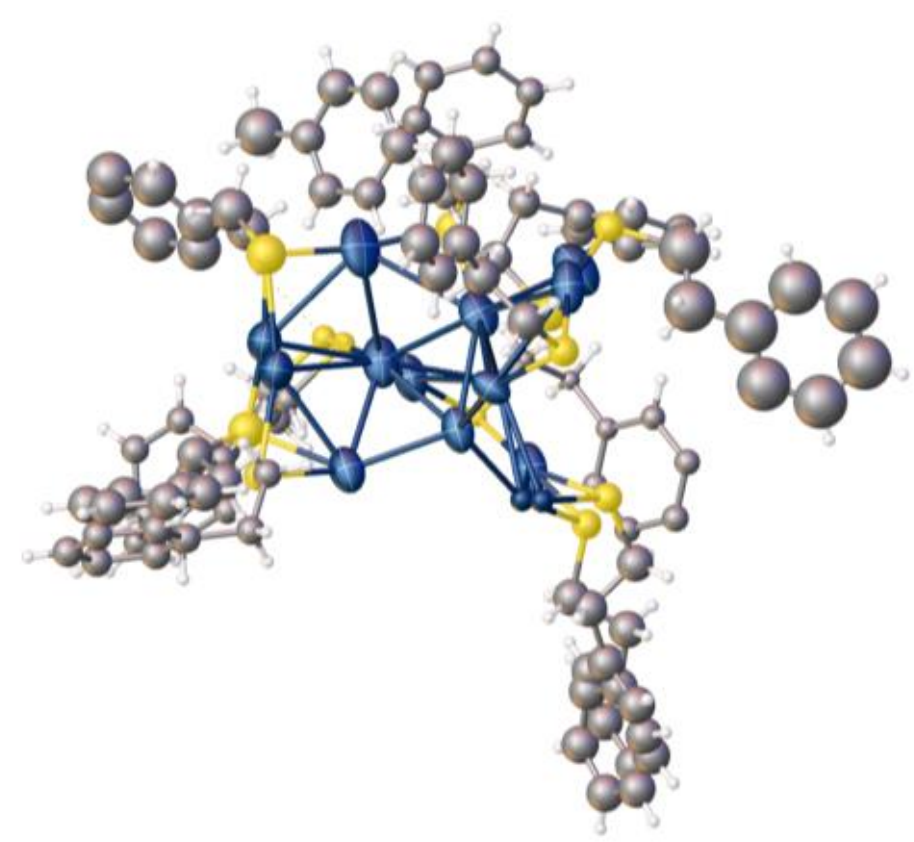

Figure S15. View of the asymmetric unit at 8.4 GPa along $\vec{a}$ with displacement parameters at 50 percent probability level.

- **Checkcif alerts and answers**

# start Validation Reply Form
_vrf_PLAT029_p08
;
PROBLEM: _diffrn_measured_fraction_theta_full value Low . 0.193 Why?
RESPONSE: These are high-pressure data taken on a home-lab diffractometer. We are limited by the use of the diamond anvil cell and the crystal system is moreover triclinic. Restrained were heavily used for the ligands to try to avoid overfitting.
;
_vrf_PLAT088_p08
;
PROBLEM: Poor Data / Parameter Ratio .................... 3.99 Note
RESPONSE: These are high-pressure data taken on a home-lab diffractometer. We are limited by the use of the diamond anvil cell and the crystal system is moreover triclinic. Restrained were heavily used for the ligands to try to avoid overfitting.
;
_vrf_PLAT342_p08
;
PROBLEM: Low Bond Precision on C-C Bonds ............... 0.12236 Ang.
RESPONSE: These are high-pressure data taken on a home-lab diffractometer. We are limited by the use of the diamond anvil cell.
;
_vrf_PLAT410_p08
;
PROBLEM: Short Intra H...H Contact H20B_4 ..H23_4 . 1.73 Ang.
RESPONSE: Although the ligand positions are partly supported by the data, since they can be somehow relocated from Fourier difference maps after removal, the precision remains limited and gets worth and worth as the pressure increases, Lig 1 and 4 may be disordered, as shown by their large displacement ellipsoids. However, due to the limited completeness on the data, we did not introduce a disordered model for them
;
_vrf_PLAT411_p08
;
PROBLEM: Short Inter H...H Contact H20B_6 ..H26_7 . 1.34 Ang.

RESPONSE: Although the ligand positions are partly supported by the data, since they can be somehow relocated from Fourier difference maps after removal, the precision remains limited and gets worth and worth as the pressure increases, Lig 1 and 4 may be disordered, as shown by their large displacement ellipsoids. However, due to the limited completeness on the data, we did not introduce a disordered model for them
;
_vrf_PLAT201_p08
;
PROBLEM: Isotropic non-H Atoms in Main Residue(s) Occ>0.5 54 Report
RESPONSE: These are high-pressure data taken on a home-lab diffractometer. We are limited by the use of the diamond anvil cell and the crystal system is moreover triclinic. Restrained were heavily used for the ligands to try to avoid overfitting. All C and S atoms were refined isotropically.
;
_vrf_PLAT241_p08
;
PROBLEM: High MainResAtom Ueq as Compared to Neighbours C20_1 Check
RESPONSE: Although the ligand positions are partly supported by the data, since they can be somehow relocated from Fourier difference maps after removal, the precision remains limited and gets worth and worth as the pressure increases, Lig 1 and 4 may be disordered , as shown by their large displacement ellipsoids. However, due to the limited completeness on the data, we did not introduce a disordered model for them
;
_vrf_PLAT242_p08
;
PROBLEM: Low MainResAtom Ueq as Compared to Neighbours S1_1 Check
RESPONSE: Although the ligand positions are partly supported by the data, since they can be somehow relocated from Fourier difference maps after removal, the precision remains limited and gets worth and worth as the pressure increases, Lig 1 and 4 may be disordered, as shown by their large displacement ellipsoids. However, due to the limited completeness on the data, we did not introduce a disordered model for them
;
_vrf_PLAT250_p08
;
PROBLEM: Large U3/U1 Ratio for <U(i,j)> Tensor(Resd 1) 4.6 Note
RESPONSE: These are high-pressure data taken on a home-lab diffractometer. We are limited by the use of the diamond anvil cell.
;
_vrf_PLAT911_p08
;
PROBLEM: Missing FCF Refl Between Thmin & STh/L= 0.600 4393 Report
RESPONSE: These are high-pressure data taken on a home-lab diffractometer. We are limited by the use of the diamond anvil cell.
;
# end Validation Reply Form

**FROM P08 TO P09**

The model from P08 was used.An alternative position for the Sulfur atom of ligand 1 was added

**P09 FINAL MODEL**

- **Crystal data and structure refinement for P09, 9.4 GPa.**

| | |
|---|---|
| Identification code | P09 |
| Empirical formula | $C_{158}H_{178}Au_{25}S_{18}$ |
| Formula weight | 7578.24 |
| Temperature/K | 299.0(4) |
| Crystal system | triclinic |
| Space group | P-1 |
| a/Å | 14.7914(8) |
| b/Å | 15.8750(11) |
| c/Å | 15.024(3) |
| α/° | 68.517(11) |
| β/° | 68.472(11) |
| γ/° | 84.796(5) |
| Volume/Å$^3$ | 3049.8(7) |
| Z | 1 |
| $\rho_{calc}$g/cm$^3$ | 4.126 |
| μ/mm$^{-1}$ | 30.298 |
| F(000) | 3389.0 |
| Crystal size/mm$^3$ | 0.183 × 0.148 × 0.029 |
| Radiation | Mo Kα (λ = 0.71073) |
| 2Θ range for data collection/° | 2.964 to 56.592 |
| Index ranges | $-15 \leq h \leq 18$, $-20 \leq k \leq 19$, $-7 \leq l \leq 7$ |
| Reflections collected | 15031 |
| Independent reflections | 2291 [$R_{int}$ = 0.0684, $R_{sigma}$ = 0.0385] |
| Data/restraints/parameters | 2291/555/576 |
| Goodness-of-fit on $F^2$ | 1.045 |
| Final R indexes [I>=2σ (I)] | $R_1$ = 0.0725, $wR_2$ = 0.1921 |
| Final R indexes [all data] | $R_1$ = 0.1004, $wR_2$ = 0.2177 |
| Largest diff. peak/hole / e Å$^{-3}$ | 2.07/-1.18 |

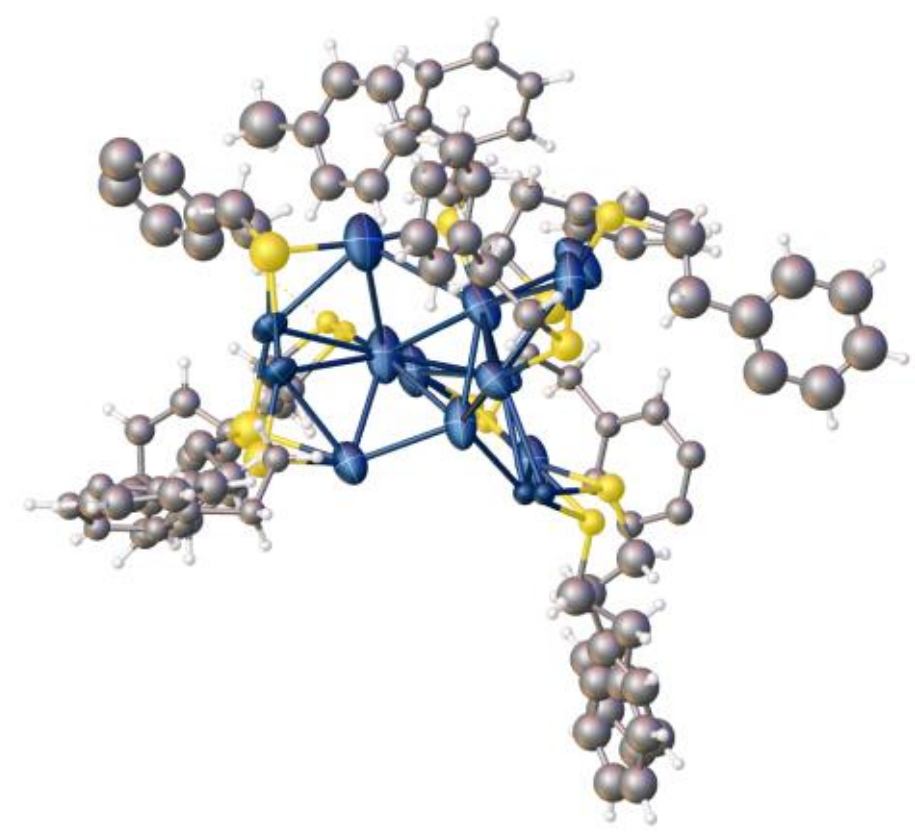

Figure S16. View of the asymmetric unit at 9.4 GPa along $\vec{a}$ with displacement parameters at 50 percent probability level.

- **Checkcif alerts and answers**

# start Validation Reply Form
_vrf_PLAT029_p09
;
PROBLEM: _diffrn_measured_fraction_theta_full value Low .     0.193 Why?
RESPONSE: These are high-pressure data taken on a home-lab diffractometer. We are limited by the use of the diamond anvil cell and the crystal system is moreover triclinic. Restrained were heavily used for the ligands to try to avoid overfitting.
;
_vrf_PLAT088_p09
;
PROBLEM: Poor Data / Parameter Ratio ....................     3.98 Note
RESPONSE: These are high-pressure data taken on a home-lab diffractometer. We are limited by the use of the diamond anvil cell and the crystal system is moreover triclinic. Restrained were heavily used for the ligands to try to avoid overfitting.
;
_vrf_PLAT342_p09
;
PROBLEM: Low Bond Precision on  C-C Bonds ...............   0.12309 Ang.
RESPONSE: These are high-pressure data taken on a home-lab diffractometer. We are limited by the use of the diamond anvil cell.
;
_vrf_PLAT410_p09
;
PROBLEM: Short Intra H...H Contact  H20B_4   ..H23_4   .      1.70 Ang.
RESPONSE: Although the ligand positions are partly supported by the data, since they can be somehow relocated  from Fourier difference maps after removal, the precision remains limited and gets worth and worth as the pressure increases, ligands  1 and6 and 4 may be disordered , as shown by their large displacement ellipsoids. However, due to the limited completeness on the data, we did not introduce a disordered model for them
;
_vrf_PLAT411_p09
;
PROBLEM: Short Inter H...H Contact  H20B_6   ..H26_7   .      1.09 Ang.

RESPONSE: Although the ligand positions are partly supported by the data, since they can be somehow relocated from Fourier difference maps after removal, the precision remains limited and gets worth and worth as the pressure increases, ligands 1 and6 and 4 may be disordered, as shown by their large displacement ellipsoids. However, due to the limited completeness on the data, we did not introduce a disordered model for them
;
_vrf_PLAT413_p09
;
PROBLEM: Short Inter XH3 .. XHn H2A ..H24_6 . 1.76 Ang.
RESPONSE: Although the ligand positions are partly supported by the data, since they can be somehow relocated from Fourier difference maps after removal, the precision remains limited and gets worth and worth as the pressure increases, ligands 1 and6 and 4 may be disordered , as shown by their large displacement ellipsoids. However, due to the limited completeness on the data, we did not introduce a disordered model for them
;
_vrf_PLAT201_p09
;
PROBLEM: Isotropic non-H Atoms in Main Residue(s) Occ>0.5 54 Report
RESPONSE: These are high-pressure data taken on a home-lab diffractometer. We are limited by the use of the diamond anvil cell and the crystal system is moreover triclinic. Restrained were heavily used for the ligands to try to avoid overfitting. All C and S atoms were refined isotropically.
;
_vrf_PLAT241_p09
;
PROBLEM: High MainResAtom Ueq as Compared to Neighbours Au12 Check
The displacement ellipsoids of the gold atoms become more elongated for several sites. This could be a possible disorder for Au12; however, this may also due to systematic errors in the data, potentially related to crystal splitting or the high pressure cell setting..
_vrf_PLAT242_p09
;
PROBLEM: Low MainResAtom Ueq as Compared to Neighbours S1_1 Check
RESPONSE: S1_1 was modelled using 2 positions but the rest of the ligand was not split due to the extremely low completeness of the data.
;
_vrf_PLAT250_p09
;
PROBLEM: Large U3/U1 Ratio for <U(i,j)> Tensor(Resd 1) 4.5 Note
The displacement ellipsoids of the gold atoms become more elongated for several sites. This may be due to systematic errors in the data, potentially related to crystal splitting or the high-pressure cell setting. This effect is highly sensitive to integration and absorption correction parameters.
;
_vrf_PLAT911_p09
;
PROBLEM: Missing FCF Refl Between Thmin & STh/L= 0.600 4363 Report
RESPONSE: These are high-pressure data taken on a home-lab diffractometer. We are limited by the use of the diamond anvil cell.
;
# end Validation Reply Form

## 3) Pressure dependence of the lattice parameters of $Au_{25}(PET)_{18}^0$

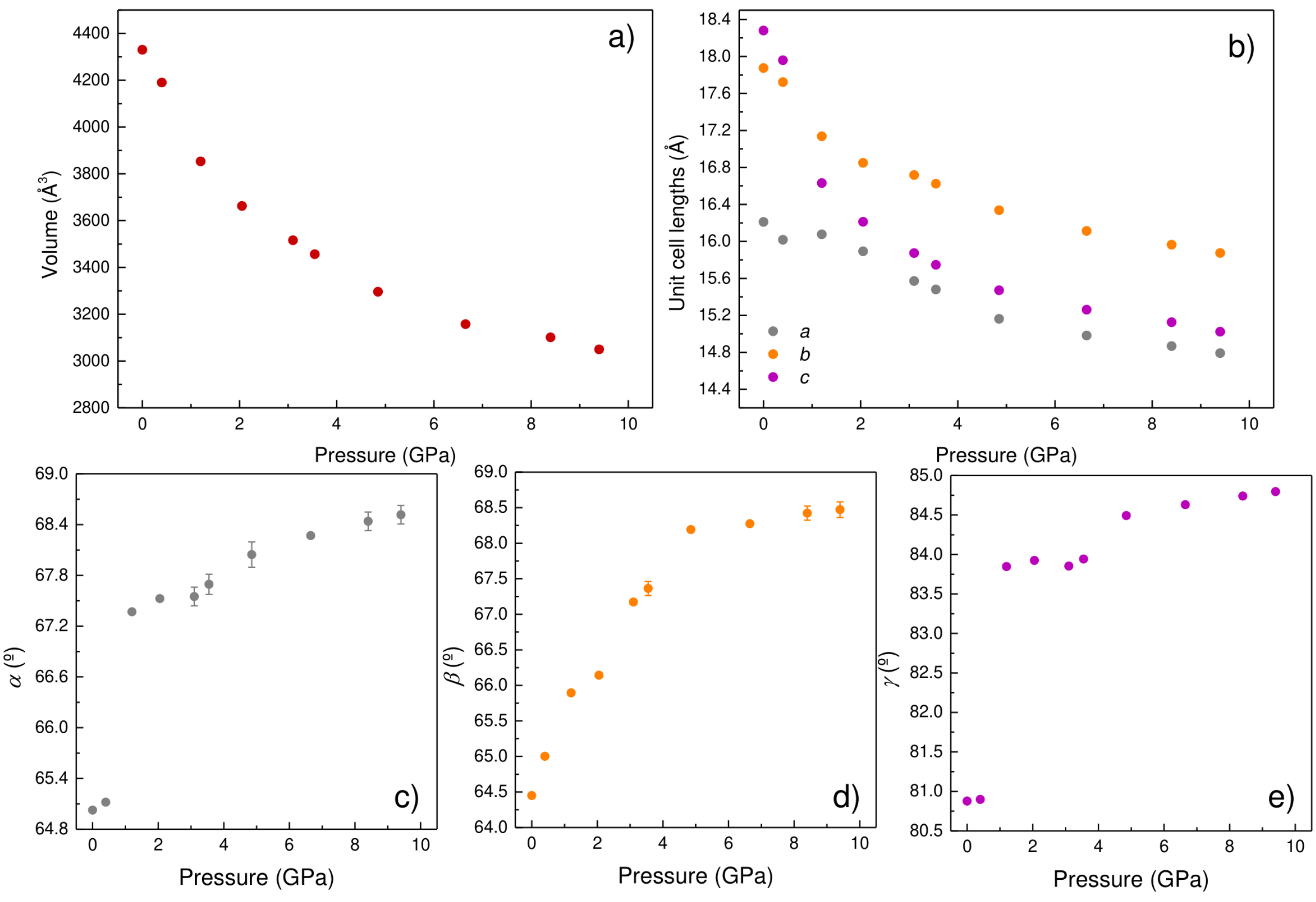

Figure S17. Pressure dependence of the lattice parameters of $Au_{25}(PET)_{18}^0$ a) Pressure dependence of the unit cell volume. b) Unit cell axes lengths as a function of pressure. c), d) and e) $\alpha$, $\beta$, and $\gamma$, respectively, as a function of pressure.

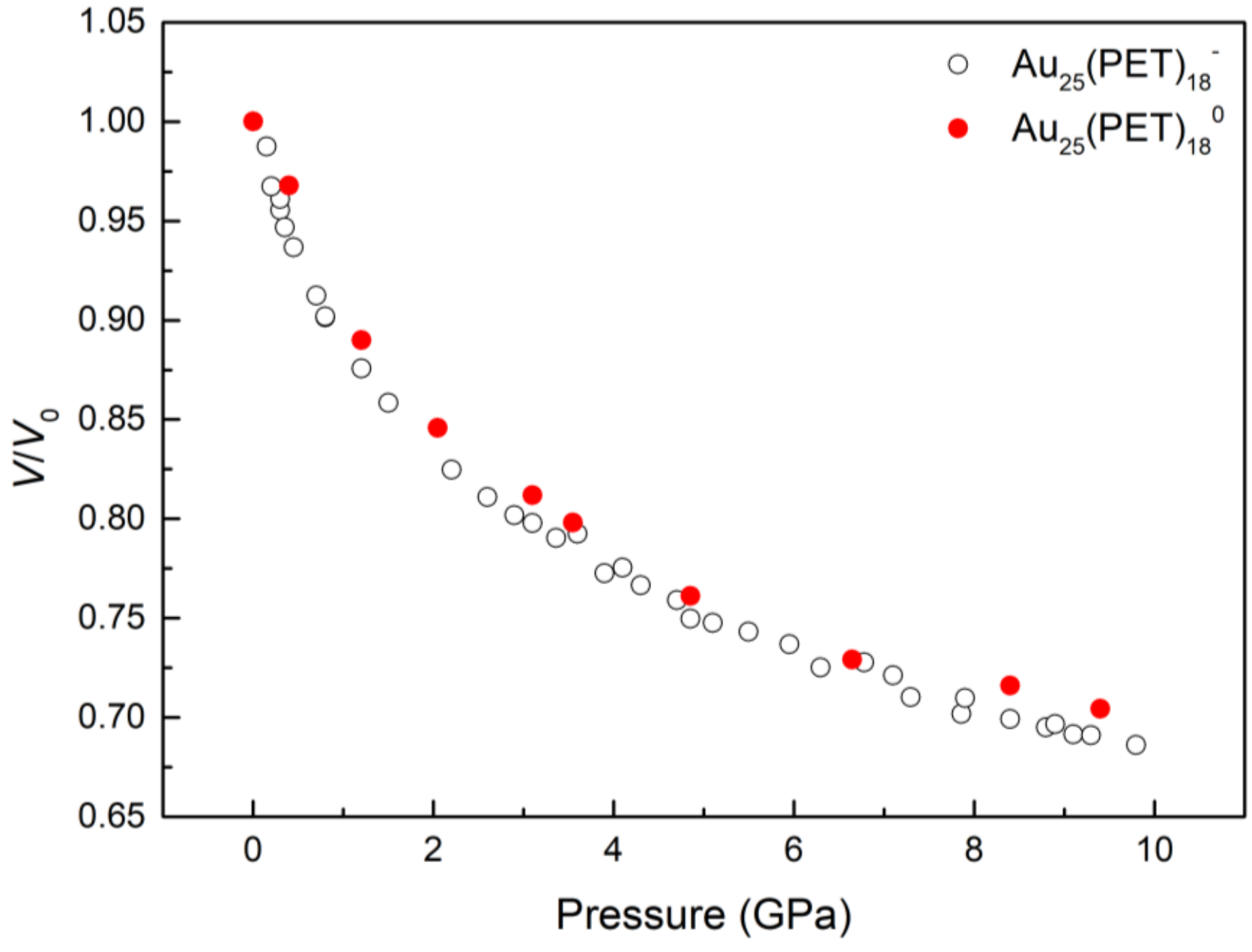

Figure S18. Pressure dependence of the relative volume of $Au_{25}(PET)_{18}^-$ (empty black dots) and $Au_{25}(PET)_{18}^0$ (filled red dots).

## 4) Pressure dependence of the atomic distances within $Au_{25}(PET)_{18}^0$

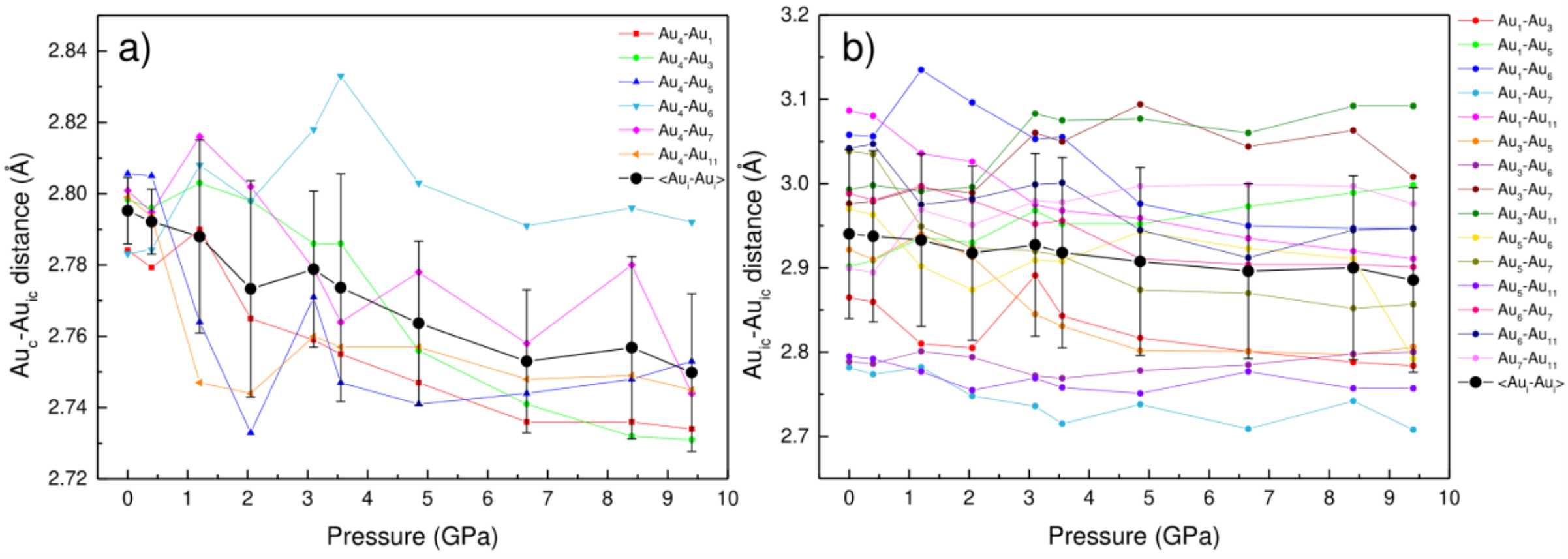

Figure S19. Evolution of individual and average interatomic distances within the $Au_{13}$ core under pressure. a) The 6 symmetry-independent radial $Au_c$-$Au_{ic}$ distances and b) the 15 symmetry-independent $Au_{ic}$-$Au_{ic}$ distances. Black circles represent the average distance for each set, with error bars corresponding to the standard deviation. Lines are guides for the eyes.

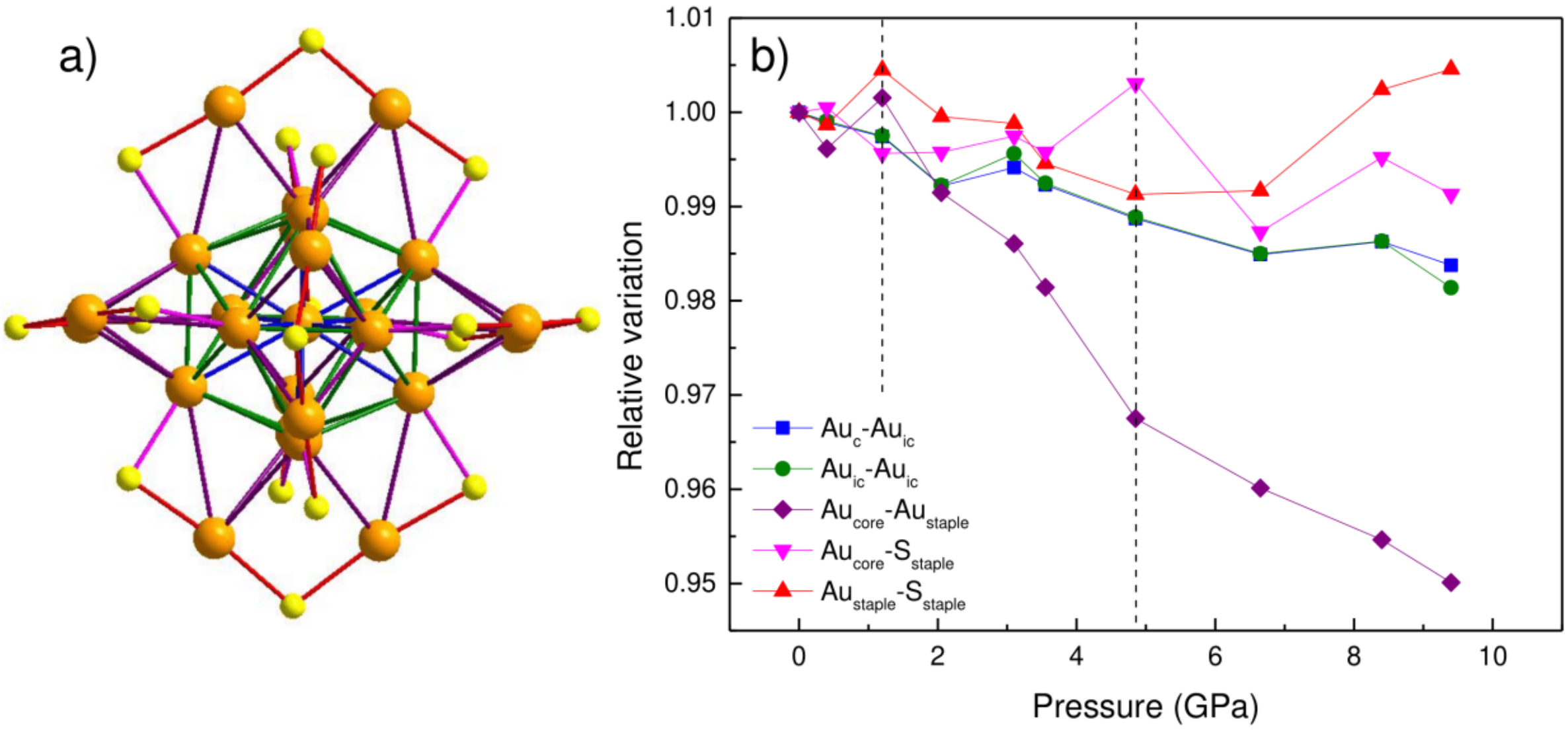

Figure S20. a) Schematic representation of the core and staples of $Au_{25}(PET)_{18}^0$. Color coding for bond types: internal $Au_c$-$Au_{ic}$ bonds (blue); $Au_{ic}$-$Au_{ic}$ bonds within the $Au_{12}$ icosahedral surface (green); $Au_{core}$-$Au_{staple}$ interfacial bonds (purple); $Au_{core}$-$Au_{staple}$ bonds (pink); and $Au_{staple}$-$S_{staple}$ bonds within the staple units (red). Gold and sulfur atoms are depicted as orange and yellow spheres, respectively. b) Pressure evolution of the average bond lengths: internal core distances ($Au_c$-$Au_{ic}$ and $Au_{ic}$-$Au_{ic}$), core-shell interface ($Au_{core}$-$Au_{staple}$ and $Au_{core}$-$S_{staple}$), and staple units ($Au_{staple}$-$S_{staple}$). Lines are guides for the eyes. Vertical dashed lines indicate the phase transition pressures. Note that above the second phase transition, the distances involving sulfur atoms should be interpreted with caution. The increased structural disorder required the use of a split-site model for the sulfur positions; consequently, the calculated distances might suffer from refinement artifacts compared to the single-site description used at lower pressures.

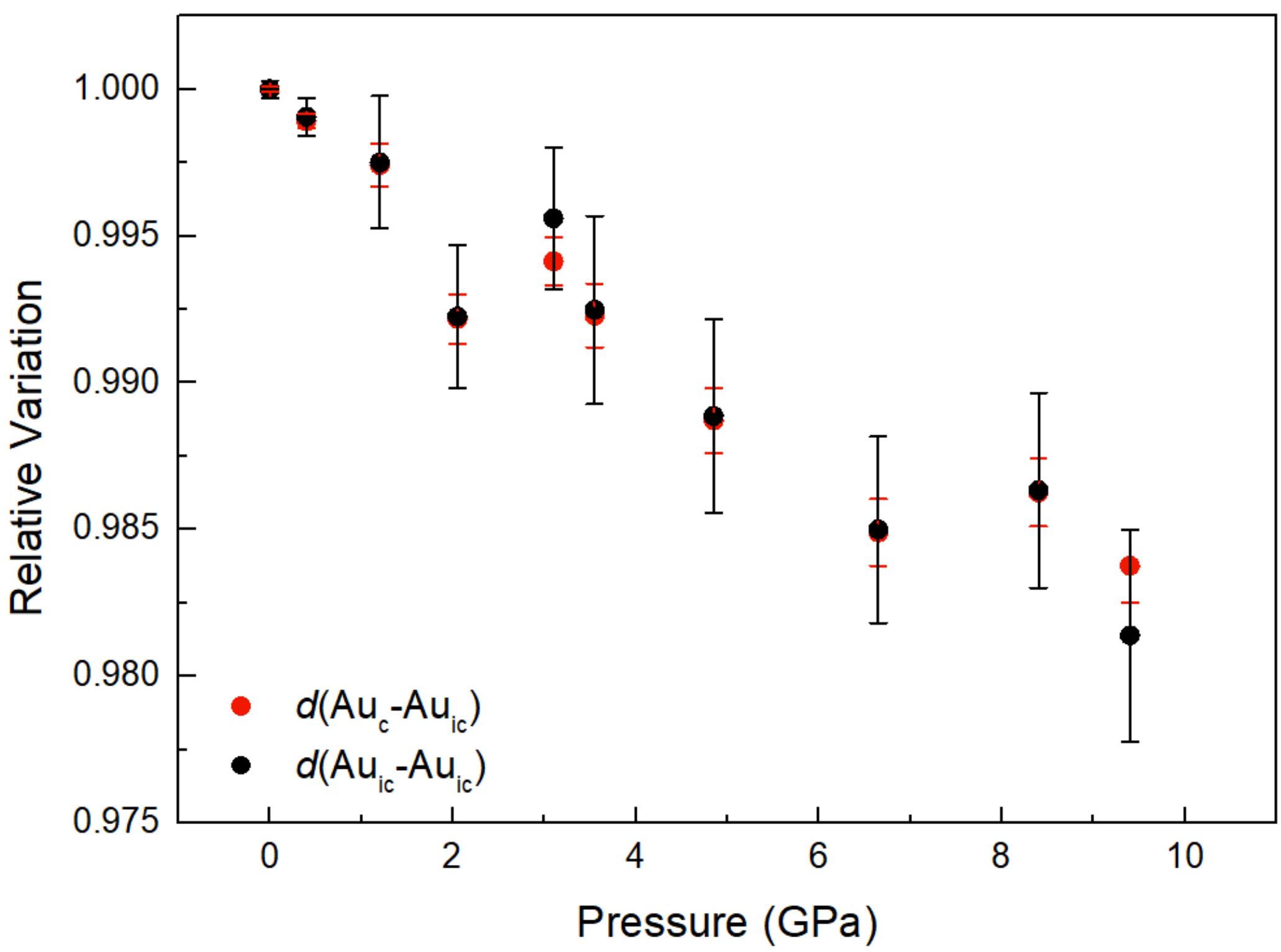

Figure S21. Pressure dependence of the relative variation of the mean radial and edge distances within the $Au_{13}$ core.

## 5) Pressure-Induced Structural Rearrangements and Ligand Shell Conformational Evolution of $Au_{25}(PET)_{18}^{0}$

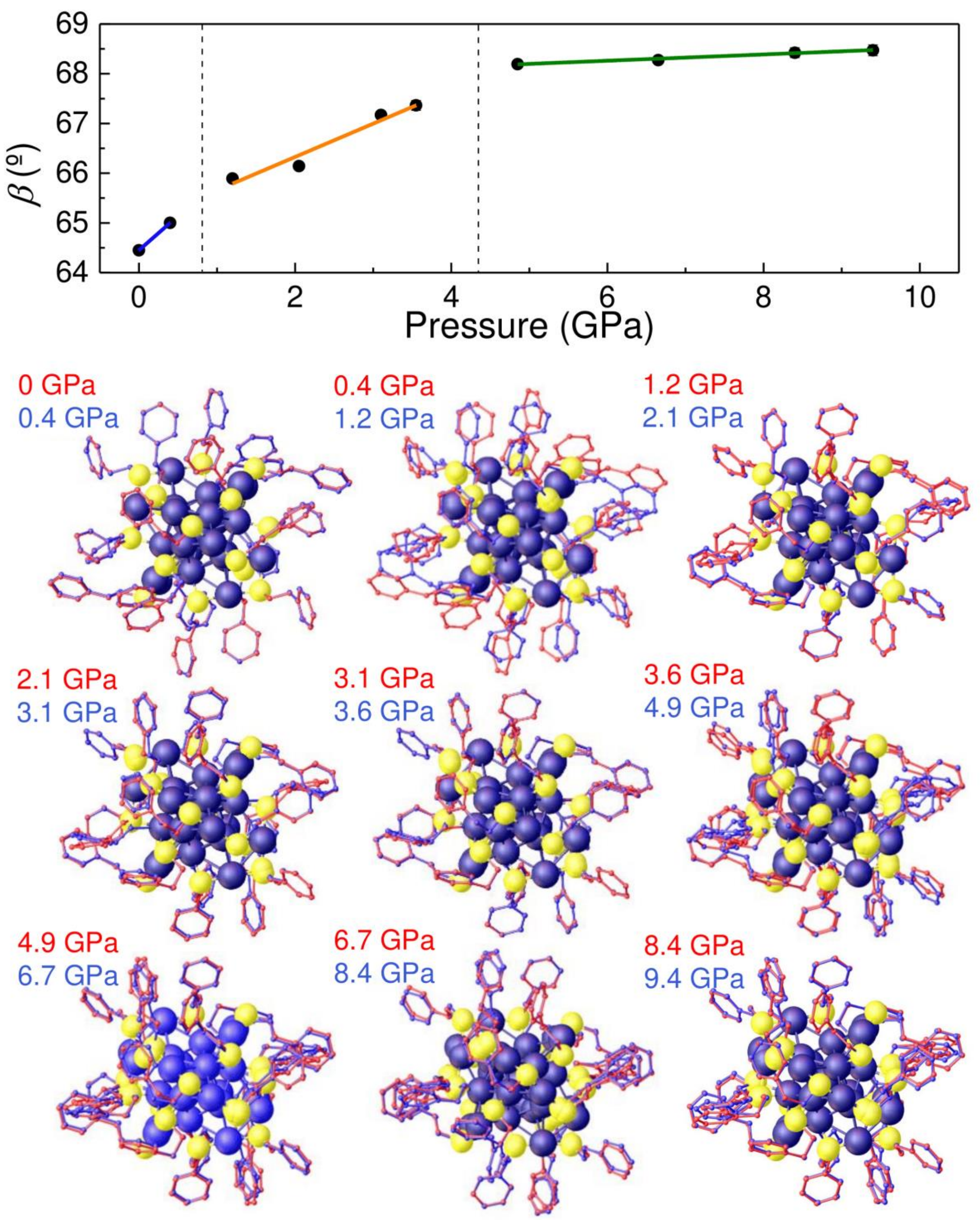

Figure S22. Structural evolutions of the structure of $Au_{25}(PET)_{18}^{0}$. (Top) Evolution of the β unit-cell parameter as a function of pressure. Solid lines represent linear fits applied to the distinct pressure regions to monitor phase transitions. (Bottom) Superposition of the structures at 2 consecutive pressures. Carbon atoms of the ligand shell at the lower pressure are shown in red, while those at the higher pressure are in blue. Gold and sulfur atoms are shown in blue and yellow, respectively. All structures were aligned using the gold atoms of the icosahedral $Au_{13}$ core to highlight the ligand reorganization.

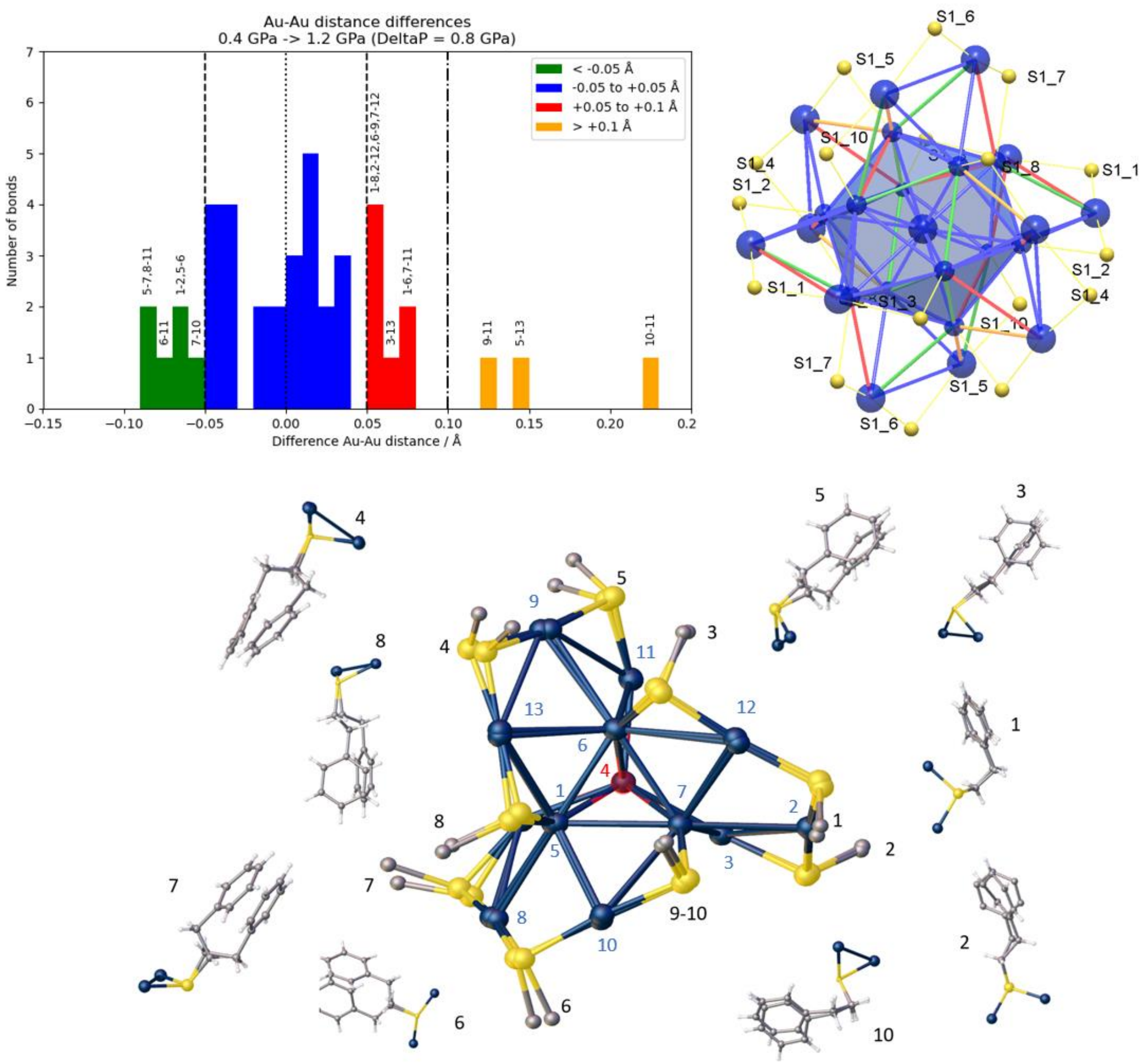

Figure S23. Structural changes associated with the first pressure-induced phase transition from 0.4 to 1.2 GPa. (Top) Differences in Au–Au bond lengths calculated from the structures at 0.4 and 1.2 GPa. The structure shown on the right is color-coded according to the bond-length differences represented in the histogram on the left. Positive and negative numbers correspond to an increase and decrease in distance, respectively. (Bottom) Variations in staple geometry. The central panel shows the superposition of one half of the cluster cores at 0.4 and 1.2 GPa; only Au atoms, S atoms, and the ligand carbon atoms bound to S are shown. The structures were aligned using the Au atoms forming the icosahedron. The central Au atom of the icosahedron is highlighted in red. Superpositions of the individual staples are also shown; these overlays were generated using the Au and S atoms of each staple. The labels correspond to the residue numbers in the CIF file.

Table S5. Aromatic interactions before the first phase transition at 0.4 GPa. Analyses of the phenyl-phenyl interaction were made in the CSD mercury software using the aromatics analyzer. This tool uses geometric description of aromatic interactions to outcome a score (from 0 to 10) indicating whether the interaction is weak (0-3), moderate (3-7) or strong (7-10). Only moderate and strong interactions were kept in the table. The number of the centroid refers to the number of the ligand in the cif file. The number 100 refers to the solvent molecule.

| Centroid1 | Centroid2 | Distance (Å) | Relative Orientation (°) | Intermolecular | Score |
|---|---|---|---|---|---|
| 1 | 3 | 4.41 | 29.89 | 1 | 6.1 |
| 1 | 3 | 5.85 | 29.89 | 1 | 6 |
| 1 | 6 | 5.59 | 64.52 | 1 | 5 |
| 1 | 10 | 5.96 | 64.37 | 1 | 4.6 |
| 1 | 4 | 6.28 | 37.81 | 1 | 4.4 |
| 1 | 2 | 6.46 | 23.81 | 1 | 4 |
| 1 | 5 | 5.66 | 79.27 | 1 | 3.7 |
| 1 | 2 | 6.46 | 23.81 | 1 | 3.3 |
| 1 | 7 | 5.68 | 84.89 | 1 | 3.1 |
| 2 | 2 | 3.69 | 0 | 1 | 9 |
| 2 | 7 | 4.92 | 61.28 | 1 | 8.9 |
| 2 | 8 | 4.71 | 60.3 | 1 | 7.9 |
| 2 | 4 | 5.57 | 36.3 | 1 | 6.2 |
| 2 | 6 | 5.06 | 80.56 | 1 | 5.7 |
| 2 | 1 | 6.46 | 23.81 | 1 | 4 |
| 2 | 1 | 6.46 | 23.81 | 1 | 3.3 |
| 3 | 6 | 4.78 | 52.04 | 1 | 8.6 |
| 3 | 3 | 4.73 | 0 | 1 | 7.9 |
| 3 | 8 | 5.57 | 20.4 | 1 | 6.3 |
| 3 | 1 | 4.41 | 29.89 | 1 | 6.1 |
| 3 | 1 | 5.85 | 29.89 | 1 | 6 |
| 3 | 4 | 6.03 | 52.03 | 1 | 4.7 |
| 3 | 10 | 6.17 | 53.15 | 1 | 3.6 |
| 4 | 5 | 4.75 | 68.69 | 1 | 8.3 |
| 4 | 100 | 5.03 | 50.12 | 1 | 8 |
| 4 | 2 | 5.57 | 36.3 | 1 | 6.2 |
| 4 | 100 | 5.26 | 50.12 | 1 | 6 |
| 4 | 10 | 5.8 | 41.87 | 1 | 5.9 |
| 4 | 3 | 6.03 | 52.03 | 1 | 4.7 |
| 4 | 1 | 6.28 | 37.81 | 1 | 4.4 |
| 4 | 4 | 6.58 | 0 | 1 | 3.5 |
| 4 | 8 | 6.3 | 43.25 | 1 | 3.5 |
| 5 | 4 | 4.75 | 68.69 | 1 | 8.3 |
| 5 | 100 | 5.33 | 62.41 | 1 | 6.6 |
| 5 | 10 | 5.27 | 37.95 | 1 | 6.4 |
| 5 | 6 | 5.62 | 50.73 | 1 | 5.2 |
| 5 | 6 | 4.4 | 50.73 | 1 | 4.8 |
| 5 | 1 | 5.66 | 79.27 | 1 | 3.7 |
| 6 | 3 | 4.78 | 52.04 | 1 | 8.6 |
| 6 | 2 | 5.06 | 80.56 | 1 | 5.7 |
| 6 | 5 | 5.62 | 50.73 | 1 | 5.2 |
| 6 | 1 | 5.59 | 64.52 | 1 | 5 |
| 6 | 5 | 4.4 | 50.73 | 1 | 4.8 |
| 6 | 8 | 6.35 | 70.08 | 1 | 4.1 |
| 7 | 8 | 4.88 | 61.17 | 1 | 8.9 |
| 7 | 2 | 4.92 | 61.28 | 1 | 8.9 |
| 7 | 10 | 5.04 | 65.2 | 1 | 8.2 |
| 7 | 100 | 4.96 | 71.76 | 1 | 6.1 |
| 7 | 8 | 5.9 | 61.17 | 1 | 3.6 |
| 7 | 1 | 5.68 | 84.89 | 1 | 3.1 |
| 8 | 7 | 4.88 | 61.17 | 1 | 8.9 |
| 8 | 10 | 4.59 | 32.79 | 1 | 8.1 |
| 8 | 2 | 4.71 | 60.3 | 1 | 7.9 |
| 8 | 8 | 5.28 | 0 | 1 | 7 |
| 8 | 3 | 5.57 | 20.4 | 1 | 6.3 |
| 8 | 6 | 6.35 | 70.08 | 1 | 4.1 |
| 8 | 7 | 5.9 | 61.17 | 1 | 3.6 |
| 8 | 4 | 6.3 | 43.25 | 1 | 3.5 |
| 10 | 7 | 5.04 | 65.2 | 1 | 8.2 |
| 10 | 8 | 4.59 | 32.79 | 1 | 8.1 |
| 10 | 5 | 5.27 | 37.95 | 1 | 6.4 |
| 10 | 4 | 5.8 | 41.87 | 1 | 5.9 |
| 10 | 1 | 5.96 | 64.37 | 1 | 4.6 |
| 10 | 100 | 5.85 | 82.74 | 1 | 4.2 |
| 10 | 3 | 6.17 | 53.15 | 1 | 3.6 |
| 100 | 4 | 5.03 | 50.12 | 1 | 8 |
| 100 | 5 | 5.33 | 62.41 | 1 | 6.6 |
| 100 | 7 | 4.96 | 71.76 | 1 | 6.1 |
| 100 | 4 | 5.26 | 50.12 | 1 | 6 |
| 100 | 10 | 5.85 | 82.74 | 1 | 4.2 |

Table S6. Aromatic interactions after the first phase transition at 1.2 GPa. Analyses of the phenyl-phenyl interaction were made in the CSD mercury software using the aromatics analyzer. This tool uses geometric description of aromatic interactions to outcome a score (from 0 to 10) indicating whether the interaction is weak (0-3), moderate (3-7) or strong (7-10). Only moderate and strong interactions were kept in the table. The number of the centroid refers to the number of the ligand in the cif file. The number 100 refers to the solvent molecule.

| Centroid1 | Centroid2 | Distance (Å) | Relative Orientation (º) | Intermolecular | Score |
|---|---|---|---|---|---|
| 1 | 5 | 4.94 | 84.37 | 1 | 8.5 |
| 1 | 3 | 5.38 | 71.23 | 1 | 7.1 |
| 1 | 7 | 6.14 | 59.7 | 1 | 4.3 |
| 1 | 10 | 6.39 | 57.25 | 1 | 3.5 |
| 2 | 8 | 4.57 | 51.02 | 1 | 8.8 |
| 2 | 4 | 5.22 | 61.35 | 1 | 7.4 |
| 2 | 7 | 4.95 | 42.06 | 1 | 7.2 |
| 2 | 2 | 5.55 | 0 | 1 | 6.9 |
| 2 | 6 | 5.48 | 52.98 | 1 | 6.6 |
| 2 | 3 | 6.57 | 66.67 | 1 | 3.3 |
| 3 | 8 | 5.1 | 54.82 | 1 | 8.5 |
| 3 | 1 | 5.38 | 71.23 | 1 | 7.1 |
| 3 | 3 | 5.38 | 0 | 1 | 6.9 |
| 3 | 5 | 5.26 | 61.03 | 1 | 5.5 |
| 3 | 2 | 6.57 | 66.67 | 1 | 3.3 |
| 4 | 2 | 5.22 | 61.35 | 1 | 7.4 |
| 4 | 100 | 5.32 | 34.32 | 1 | 7.1 |
| 4 | 8 | 5.68 | 14.28 | 1 | 6.1 |
| 4 | 100 | 5.1 | 34.32 | 1 | 5.4 |
| 4 | 5 | 5.87 | 71.85 | 1 | 4.9 |
| 5 | 6 | 5.02 | 75.99 | 1 | 8.6 |
| 5 | 1 | 4.94 | 84.37 | 1 | 8.5 |
| 5 | 6 | 5.28 | 75.99 | 1 | 5.9 |
| 5 | 3 | 5.26 | 61.03 | 1 | 5.5 |
| 5 | 4 | 5.87 | 71.85 | 1 | 4.9 |
| 5 | 9 | 6.23 | 89.24 | 1 | 3.8 |
| 5 | 5 | 6.52 | 0 | 1 | 3.3 |
| 6 | 5 | 5.02 | 75.99 | 1 | 8.6 |
| 6 | 100 | 4.87 | 56.89 | 1 | 8.2 |
| 6 | 2 | 5.48 | 52.98 | 1 | 6.6 |
| 6 | 5 | 5.28 | 75.99 | 1 | 5.9 |
| 6 | 7 | 6.46 | 50.8 | 1 | 4.1 |
| 6 | 1 | 6.43 | 35.75 | 1 | 3.5 |
| 7 | 100 | 4.82 | 71.47 | 1 | 8.6 |
| 7 | 2 | 4.95 | 42.06 | 1 | 7.2 |
| 7 | 10 | 5.05 | 39.18 | 1 | 6.8 |
| 7 | 8 | 5.54 | 87.12 | 1 | 6.2 |
| 7 | 9 | 6.21 | 30.08 | 1 | 5.1 |
| 7 | 1 | 6.14 | 59.7 | 1 | 4.3 |
| 7 | 6 | 6.46 | 50.8 | 1 | 4.1 |
| 7 | 9 | 5.7 | 30.08 | 1 | 4.1 |
| 7 | 8 | 5.87 | 87.12 | 1 | 3.4 |
| 8 | 2 | 4.57 | 51.02 | 1 | 8.8 |
| 8 | 3 | 5.1 | 54.82 | 1 | 8.5 |
| 8 | 7 | 5.54 | 87.12 | 1 | 6.2 |
| 8 | 4 | 5.68 | 14.28 | 1 | 6.1 |
| 8 | 8 | 5.58 | 0 | 1 | 3.9 |
| 8 | 7 | 5.87 | 87.12 | 1 | 3.4 |
| 9 | 10 | 5.48 | 9.83 | 1 | 6.7 |
| 9 | 7 | 6.21 | 30.08 | 1 | 5.1 |
| 9 | 7 | 5.7 | 30.08 | 1 | 4.1 |
| 9 | 5 | 6.23 | 89.24 | 1 | 3.8 |
| 9 | 100 | 5.6 | 72.48 | 1 | 3.1 |
| 10 | 7 | 5.05 | 39.18 | 1 | 6.8 |
| 10 | 9 | 5.48 | 9.83 | 1 | 6.7 |
| 10 | 4 | 5.54 | 76.17 | 1 | 6 |
| 10 | 3 | 6.24 | 52.93 | 1 | 3.6 |
| 10 | 1 | 6.39 | 57.25 | 1 | 3.5 |

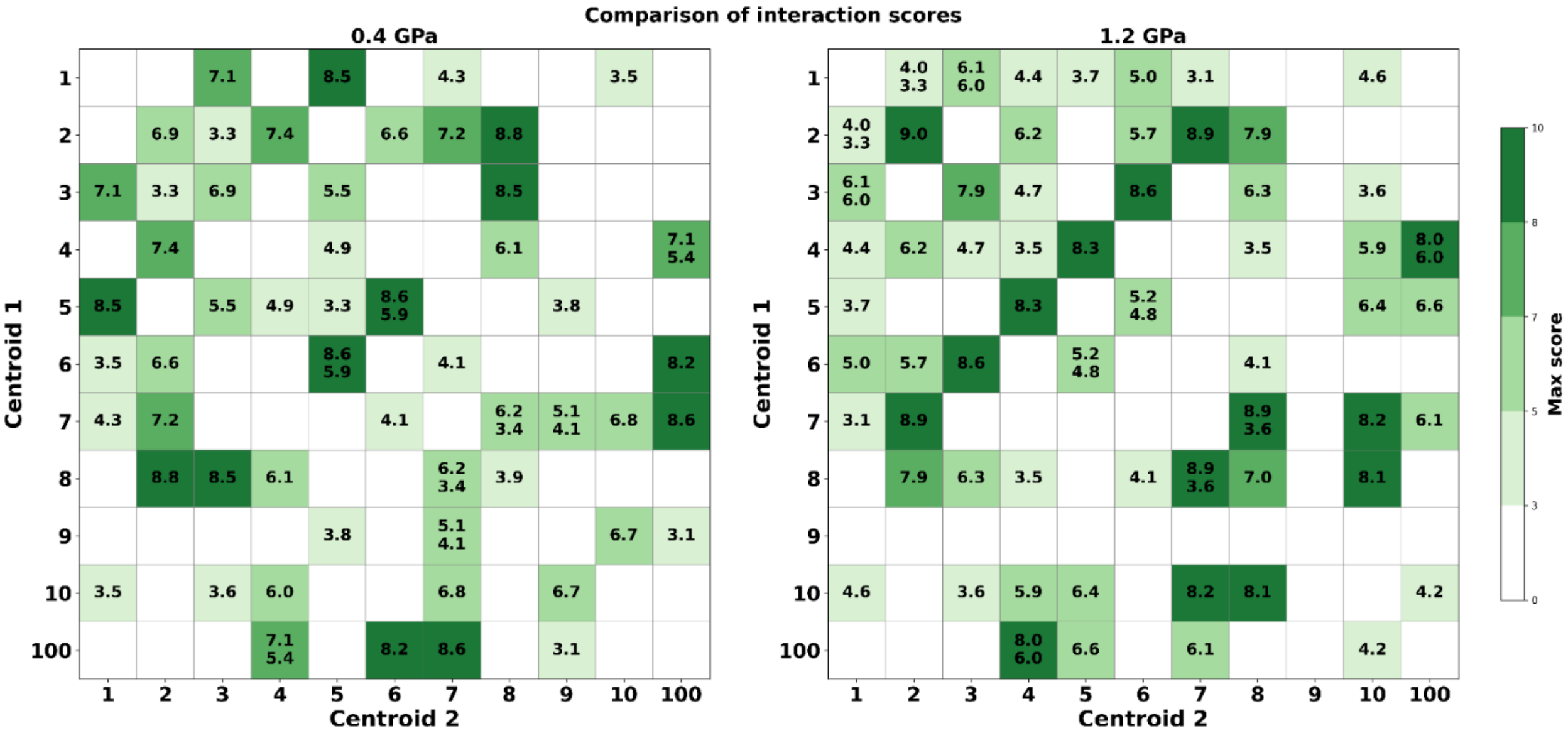

Figure S24. Visual comparison of the scores obtained by the aromatics analyzer (see table S6 and S7) before and after the first phase transition. Only moderate (score 3-7) or strong (score 7-10) were kept. The number of the centroid refers to the number of the ligand in the cif file. The number 100 refers to the solvent molecule. Ligand 9 and 10 at 0.4 GPa represents 2 components of a disordered structure. This disorder disappears after the phase transition and only ligand 10 is present at 1.2 GPa. Ligand 9 was kept to keep the matrix size identical, allowing a quicker visual comparison.

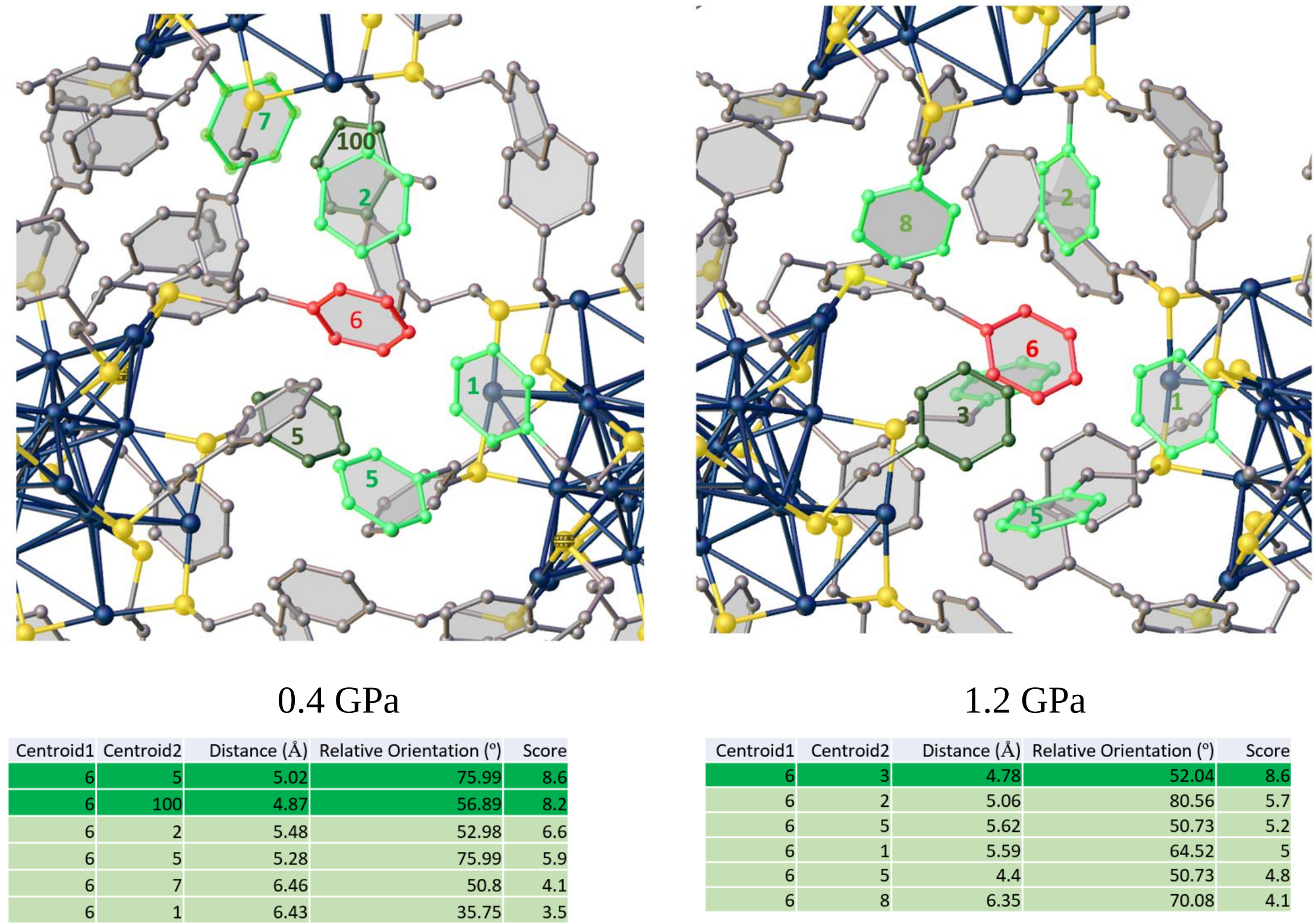

0.4 GPa

| Centroid1 | Centroid2 | Distance (Å) | Relative Orientation (°) | Score |
|---|---|---|---|---|
| 6 | 5 | 5.02 | 75.99 | 8.6 |
| 6 | 100 | 4.87 | 56.89 | 8.2 |
| 6 | 2 | 5.48 | 52.98 | 6.6 |
| 6 | 5 | 5.28 | 75.99 | 5.9 |
| 6 | 7 | 6.46 | 50.8 | 4.1 |
| 6 | 1 | 6.43 | 35.75 | 3.5 |

1.2 GPa

| Centroid1 | Centroid2 | Distance (Å) | Relative Orientation (°) | Score |
|---|---|---|---|---|
| 6 | 3 | 4.78 | 52.04 | 8.6 |
| 6 | 2 | 5.06 | 80.56 | 5.7 |
| 6 | 5 | 5.62 | 50.73 | 5.2 |
| 6 | 1 | 5.59 | 64.52 | 5 |
| 6 | 5 | 4.4 | 50.73 | 4.8 |
| 6 | 8 | 6.35 | 70.08 | 4.1 |

Figure S25. View of the aromatic interactions calculated for ligand 6 with the numbering of the ligands. View along the *a*-axis. The number of the centroid refers to the number of the ligand in the cif file. The number 100 refers to the solvent molecule. Only moderate (score 3-7, light green) or strong (score 7-10, dark green) interactions, as defined by the CSD mercury aromatics analyser, are displayed.

Table S7. Geometrical parameters describing the staple geometry from 0 to 3.6 GPa

| Parameter | 0 GPa | 0.4 GPa | 1.2 GPa | 2.1 GPa | 3.1 GPa | 3.6 GPa |
|---|---|---|---|---|---|---|
| **Au12-S1_1** | 2.306 | 2.317 | 2.31 | 2.3 | 2.33 | 2.33 |
| **Au2-S1_1** | 2.302 | 2.282 | 2.31 | 2.25 | 2.24 | 2.24 |
| **Au2-S1_1-Au12** | 100.14 | 100.34 | 102 | 102.8 | 101.2 | 100.1 |
| **Au2-S1_1-C20_1-C21_1** | 82.8 | 85.5 | 89 | 84 | 85 | 83 |
| **C20_1-C21_1-C22_1-C23_1** | 118.3 | 120.8 | 114 | 111 | 118 | 117 |
| **C20_1-C21_1-C22_1-C27_1** | -63.4 | -62 | -67 | -69 | -67 | -65 |
| **S1_1-C20_1-C21_1-C22_1** | -62.7 | -66 | -67 | -59 | -62 | -61 |
| **Au2-S1_2** | 2.3 | 2.308 | 2.31 | 2.3 | 2.37 | 2.45 |
| **Au2-S1_2-Au3** | 89.25 | 88.7 | 88.4 | 88.9 | 86.7 | 84.1 |
| **Au2-S1_2-C20_2-C21_2** | 68.2 | 65.6 | 85 | 86 | 77 | 76 |
| **Au3-S1_2** | 2.377 | 2.38 | 2.366 | 2.355 | 2.32 | 2.35 |
| **C20_2-C21_2-C22_2-C23_2** | 78.4 | 80 | 106 | 106 | 130 | 131 |
| **C20_2-C21_2-C22_2-C27_2** | -97.3 | -97 | -76 | -75 | -51 | -50 |
| **S1_2-C20_2-C21_2-C22_2** | 170.3 | 171.5 | 176 | 173 | 159 | 159 |
| **Au12-S1_3** | 2.309 | 2.315 | 2.31 | 2.31 | 2.35 | 2.35 |
| **Au6-S1_3** | 2.359 | 2.365 | 2.34 | 2.37 | 2.39 | 2.38 |
| **Au6-S1_3-Au12** | 90.46 | 89.7 | 89.3 | 87.8 | 87.4 | 87.2 |

| **Au6-S1_3-C20_3-C21_3** | -153.7 | -152 | -160 | -150 | -146 | -149 |
|---|---|---|---|---|---|---|
| **C20_3-C21_3-C22_3-C23_3** | 159.8 | 157 | 79 | 72 | 61 | 67 |
| **C20_3-C21_3-C22_3-C27_3** | -24 | -27 | -102 | -108 | -119 | -113 |
| **S1_3-C20_3-C21_3-C22_3** | -176.7 | -175 | -161 | -164 | -162 | -168 |
| **Au13-S1_4** | 2.293 | 2.29 | 2.302 | 2.275 | 2.298 | 2.36 |
| **Au9-S1_4** | 2.302 | 2.293 | 2.32 | 2.35 | 2.05 | 1.8 |
| **Au9-S1_4-Au13** | 101.17 | 101.3 | 92.8 | 90.7 | 100.7 | 106 |
| **Au9-S1_4-C20_4-C21_4** | 61 | 59 | -58 | -62 | -59 | -57 |
| **C20_4-C21_4-C22_4-C23_4** | 101.9 | 103 | -150 | 31 | 32 | 33 |
| **C20_4-C21_4-C22_4-C27_4** | -76.9 | -73 | 32 | -150 | -152 | -148 |
| **S1_4-C20_4-C21_4-C22_4** | 174.4 | 176 | 160 | 164 | 156 | 154 |
| **Au11-S1_5** | 2.36 | 2.361 | 2.32 | 2.32 | 2.37 | 2.37 |
| **Au9-S1_5** | 2.311 | 2.29 | 2.37 | 2.29 | 2.44 | 2.4 |
| **Au9-S1_5-Au11** | 89.54 | 89.6 | 93 | 94.1 | 88.9 | 88.3 |
| **Au9-S1_5-C20_5-C21_5** | -39.6 | -45 | 92 | 146 | 139 | 144 |
| **C20_5-C21_5-C22_5-C23_5** | 96.8 | 100 | 18 | 139 | 114 | 102 |
| **C20_5-C21_5-C22_5-C27_5** | -81 | -75 | -163 | -44 | -69 | -80 |
| **S1_5-C20_5-C21_5-C22_5** | 174.7 | 175 | 167 | -175 | -135 | -131 |
| **Au10-S1_6** | 2.299 | 2.304 | 2.286 | 2.31 | 2.286 | 2.28 |
| **Au8-S1_6** | 2.296 | 2.299 | 2.31 | 2.33 | 2.33 | 2.39 |
| **Au8-S1_6-Au10** | 99.7 | 99.4 | 100.1 | 98.1 | 97.9 | 95.5 |
| **Au8-S1_6-C20_6-C21_6** | 166.3 | 164.5 | -37 | -37 | -24 | -25 |
| **C20_6-C21_6-C22_6-C23_6** | 124.2 | 123 | 53 | 52 | 31 | 32 |
| **C20_6-C21_6-C22_6-C27_6** | -58.1 | -58 | -124 | -129 | -150 | -148 |
| **S1_6-C20_6-C21_6-C22_6** | -176.1 | -177 | 166 | 171 | 170 | 172 |
| **Au1-S1_7** | 2.36 | 2.357 | 2.362 | 2.335 | 2.334 | 2.37 |
| **Au1-S1_7-Au8** | 87.01 | 86.37 | 87.9 | 86.9 | 85.1 | 84.3 |
| **Au1-S1_7-C20_7-C21_7** | 129.6 | 134 | -53 | -45 | -42 | -49 |
| **Au8-S1_7** | 2.294 | 2.296 | 2.31 | 2.32 | 2.33 | 2.31 |
| **C20_7-C21_7-C22_7-C23_7** | 90.6 | 86 | -67 | -77 | -74 | -68 |
| **C20_7-C21_7-C22_7-C27_7** | -90.3 | -95 | 112 | 103 | 103 | 110 |
| **S1_7-C20_7-C21_7-C22_7** | 179.3 | 179.5 | -176 | -174 | -178 | -172 |
| **Au13-S1_8** | 2.3 | 2.299 | 2.292 | 2.29 | 2.273 | 2.25 |
| **Au5-S1_8** | 2.373 | 2.37 | 2.38 | 2.35 | 2.33 | 2.26 |
| **Au5-S1_8-Au13** | 86.61 | 86 | 91.2 | 91.4 | 89.8 | 92 |
| **Au5-S1_8-C20_8-C21_8** | -57.1 | -54 | -46 | -45 | -46 | -49 |
| **C20_8-C21_8-C22_8-C23_8** | 61 | 59 | 109 | 110 | 111 | 113 |
| **C20_8-C21_8-C22_8-C27_8** | -120 | -120 | -68 | -69 | -68 | -65 |
| **S1_8-C20_8-C21_8-C22_8** | -176.8 | 177 | -176 | -178 | -174 | -176 |
| **Au10-S1_9** | 2.319 | 2.301 | | | | |
| **Au7-S1_9** | 2.361 | 2.364 | | | | |
| **Au7-S1_9-Au10** | 91.2 | 91.22 | | | | |
| **Au7-S1_9-C20_9-C21_9** | -170 | -171 | | | | |
| **C20_9-C21_9-C22_9-C23_9** | 115 | 119 | | | | |
| **C20_9-C21_9-C22_9-C27_9** | -62 | -58 | | | | |
| **S1_9-C20_9-C21_9-C22_9** | -61 | -60 | | | | |

| **Au10-S1_10** | 2.319 | 2.301 | 2.325 | 2.293 | 2.301 | 2.321 |
|---|---|---|---|---|---|---|
| **Au7-S1_10** | 2.361 | 2.364 | 2.36 | 2.4 | 2.36 | 2.4 |
| **Au7-S1_10-Au10** | 91.2 | 91.22 | 88.7 | 86.9 | 89.9 | 88.8 |
| **Au7-S1_10-C20_10-C21_10** | 156 | 155 | 161 | 159 | 164 | 165 |
| **C20_10-C21_10-C22_10-C23_10** | 49 | 54 | 71 | 75 | 76 | 80 |
| **C20_10-C21_10-C22_10-C27_10** | -125 | -124 | -108 | -104 | -103 | -98 |
| **S1_10-C20_10-C21_10-C22_10** | 51 | 49 | 50 | 53 | 44 | 40 |

## 6) Computational Assessment of Structural Centrosymmetry of $Au_{25}(PET)_{18}^{0}$ under Pressure

Periodic DFT calculations were performed to evaluate the probability of maintaining inversion symmetry in $Au_{25}(PET)_{18}^{0}$ under pressure. A model structure was designed from the resolved experimental X-ray structure and optimized with no symmetry constraints for applied pressures ranging from 0 to 12 GPa. The computations were carried out with the plane-wave-based *Quantum Espresso 7.3.1* program package [S1,S2] utilizing the dispersion-corrected PBE-D3 functional [S3,S4] and pseudopotentials obtained from the Standard Solid-State Pseudopotentials (SSSP) PBE Efficiency 1.3.0 library [S5]. The wavefunctions and charge density cutoffs were expanded in a planewave basis set with cutoffs of 60 Ry and 480 Ry, respectively. For the study of this large molecular crystal, the sampling of the Brillouin zone was restricted to the Γ point and a 5 mRy Gaussian smearing was used for the electronic occupations. Finally, the optimized structures at each pressure point were analyzed for additional symmetry and inversion probability using the *PLATON* software package [S6,S7]. Table S8 summarizes the calculated probability of the structure remaining centrosymmetric as a function of the applied pressure.

Table S8. Calculated probability of centrosymmetry for $Au_{25}(PET)_{18}^{0}$ at different pressures. Fit between the optimized structure and a centrosymmetric structure for the different pressures tested, as calculated by the ADDSYM algorithm in PLATON. *Below 70%, PLATON does not give indication of a missed additional symmetry.

| Pressure (GPa) | Centrosymmetry Probability (%) |
|---|---|
| 2 | 92 |
| 4 | 91 |
| 8 | 82 |
| 10 | 74 |
| 12 | less than 70* |